\documentclass[useAMS,usenatbib]{mnras}
\usepackage{graphicx,amsmath,color,amssymb}
\usepackage[svgnames]{xcolor}
\usepackage{paralist}
\usepackage{mathtools}
\usepackage{hyperref}

\newcommand\colden{\ensuremath{N_{\text{HI}}}}
\newcommand{\Lya}{Lyman-$\alpha$}

\newcommand{\commentout}[1]{}

\newcommand{\Data}{\mathcal{D}}

\newcommand{\gp}{\textsc{gp}}

\newcommand{\normal}{\mathcal{N}}

\newcommand{\lambdarest}{\lambda_{\textrm{rest}}}

\newcommand{\lambdaobs}{\lambda_{\textrm{obs}}}

\newcommand{\qso}{\textsc{qso}}

\newcommand{\Kvec}{\boldsymbol{K}}

\newcommand{\AAtext}{\textrm{\AA}}

\newcommand{\Yvec}{\boldsymbol{Y}}
\newcommand{\Mvec}{\boldsymbol{M}}

\newcommand{\lognhi}{\log_{10}{N_{\textrm{HI}}}}

\newcommand{\zdla}{z_{\textrm{DLA}}}

\newcommand\zqso{{\ensuremath{z_{\text{QSO}}}}}
\newcommand\DLA{\ensuremath{\text{D}^1}}
\newcommand\noDLA{\ensuremath{\text{D}^0}}

\newcommand{\mzqso}{\noDLA}
\newcommand{\rszqso}{\mathrm{rs}_{\zqso}}

\newcommand{\yobs}{x}
\newcommand{\zmap}{z_\mathrm{MAP}}

\newcommand{\murszqso}{\mu \circ \rszqso}
\newcommand{\Krszqso}{K \circ \rszqso}
\newcommand{\Knoise}{K_N}
\newcommand{\obsflux}{x}
\newcommand{\emissiony}{\tilde{y}}

\newcommand{\xnorm}{\bar{x}} 
\newcommand{\uniform}{\mathcal{U}}
\newcommand{\zqsomin}{\zqso{_\mathrm{min}}}
\newcommand{\zqsomax}{\zqso{_\mathrm{max}}}

\newcommand\redshift{\ensuremath{\text{rs}}}
\newcommand\rf[1]{\tilde{#1}}
\newcommand\rfdla[1]{\check{#1}}
\newcommand\obs{\ensuremath{x}}
\newcommand\comp[2]{\ensuremath{{#1} \!\circ\! {#2}}}

\title[Automated Quasar Redshifts]{Automated Measurement of Quasar Redshift with a Gaussian Process}

\author[]{Leah Fauber$^1$\thanks{E-mail: jfaub001@ucr.edu}, Ming-Feng Ho$^1$, Simeon Bird$^1$, Christian R. Shelton$^1$, \newauthor
Roman Garnett$^2$, Ishita Korde$^1$ \vspace{1.5mm}\\
$^1$ University of California Riverside, Riverside, CA\\
$^2$ Washington University in St.\ Louis, One Brookings Drive, St.\ Louis, MO 63130, USA}
\begin{document}
\label{firstpage}
\pagerange{\pageref{firstpage}--\pageref{lastpage}}
\maketitle

\begin{abstract}
We develop an automated technique to measure quasar redshifts in the Baryon
	Oscillation Spectroscopic Survey (BOSS) of the Sloan Digital Sky Survey (SDSS).
	Our technique is an extension of an earlier Gaussian process method for
	detecting damped Lyman-$\alpha$ absorbers (DLAs) in quasar spectra with known
	redshifts. We apply this technique to a subsample of SDSS DR12 with BAL quasars removed and redshift larger than $2.15$.
	We show that we are broadly competitive to existing quasar redshift
	estimators, disagreeing with the PCA redshift by more than $0.5$ in only $0.38\%$ of spectra. Our method produces a probabilistic density function
	for the quasar redshift, allowing quasar redshift uncertainty to be propagated to downstream users. We apply this method to detecting DLAs, accounting in a Bayesian fashion for redshift uncertainty. Compared to our earlier method with a known quasar redshift, we have a moderate decrease in our ability to detect DLAs, predominantly in the noisiest spectra. The area under curve drops from $0.96$ to $0.91$. Our code is publicly available.
\end{abstract}


%
%
%

\section{Introduction}

Estimating redshifts using spectroscopy is a well-explored technique in astronomy. Spectroscopy uses the presence of lines at known emission wavelengths to estimate the redshift of an object. While quasi-stellar objects (QSOs, or quasars) contain multiple strong emission lines, the presence of quasar outflows mean that these lines often have an intrinsic Doppler shift from their rest positions, leading to hard to quantify redshift errors \citep{Gaskell:1982, Shen:2016}. The Sloan Digital Sky Survey (SDSS) \citep{Eisenstein:2011, Dawson:2013, Alam:2015} presents a further challenge due to the low signal to noise of many of the spectra. Redshift estimation in Data Release 14 (DR14Q) is done using $4$ different techniques. These include principal component analysis (PCA) using DR5 as a training sample \citep{Hewett:2010, Schneider:2010}, automated fitting to the MgII emission line, and a partial visual inspection survey \citep{paris2018sloan}. Techniques differ, on average, by around $100$ km/s, with a velocity dispersion of $\sim 500$ km/s. Furthermore, they fail to converge for about $0.5\%$ of objects. Estimation of quasar redshift, $\zqso$, must be accurate to achieve the scientific goals of spectroscopic surveys. Systematic and statistical errors in redshift estimation reduce the strength of the Baryon Acoustic Oscillation (BAO) signal \citep{Dawson:2016}.

Each new generation of spectroscopic survey roughly doubles the number of quasar spectra, such that DR14Q contains $1.8\times 10^5$ quasars with Lyman-$\alpha$ absorption in the Baryon Oscillation Sky Survey (BOSS) \citep{paris2017sloan}. The next generation Dark Energy Spectroscopic Instrument (DESI) will ultimately contain $7\times 10^5$ Lyman-$\alpha$ quasars \citep{DESI}. Algorithmic inspection of quasar spectra, already essential, will become yet more necessary to keep pace with data collection.

We estimate quasar redshifts using a Gaussian process (GP) model for quasar
spectra. Compared to existing redshift estimation techniques, our model is conceptually most similar to PCA redshifts, although we improve on them by explicitly accounting for noise in the spectrum. All emission lines in the redshift range are fit simultaneously. Our model uses the existing catalogue as a prior to constrain the expected offsets of each line from the intrinsic emission redshift. In principle, we are also able to learn correlations between emission line width and velocity offset \citep{Mason:2017}.

We build on the work of \cite{Garnett:2016, Bird:2017, Ho:2020}. \cite{Garnett:2016} built a GP model for quasar spectra and combined it with an analytic Voigt profile to find Damped Lyman-$\alpha$ absorbers (DLAs), strong neutral hydrogen absorption lines corresponding to the gas surrounding high redshift dwarf galaxies \citep{Wolfe:1986, Prochaska:1997, Haehnelt:1998, Bird:2014}. We extend the emission model implicit in that work to the whole quasar spectrum between $3000$ and $910$\AA. We then try to use all information about the shape and properties of the quasar to estimate the quasar redshift. In practice redshift estimation in our model is driven by the fit to well-known emission peaks, especially MgII, CIII, CIV and Lyman-$\alpha$. We train the model using the SDSS pipeline quasar redshift estimate, and use the trained model to estimate the redshift of quasars outside the training set. To verify our method, we check our derived quasar redshifts against the other redshift estimates included in the SDSS catalogue and show that they are competitive to other techniques.

We also provide a modified DLA catalogue for SDSS DR12 to demonstrate that we can detect DLAs while marginalizing out redshift uncertainty. To validate the results, we compare them to catalogues from the template fitting code of \cite{noterdaeme2012column}, the SDSS visual inspection survey and the neural network based model of \cite{parks2018deep}. We require three separate catalogues in order to generate and thus compare to a ``best two of three'' catalogue to serve as ground truth. We emphasise that we use visual inspection, an non-automated technique which will not be available for future data releases, only for validation.

Section~\ref{sec:gpmodel} defines our overall emission model, Section~\ref{sec:pure_z} describes our redshift estimation, and Section~\ref{sec:dlafind} describes DLA finding. We summarize the main results from \cite{Garnett:2016}, on which our model is heavily based and point explicitly to changes. Our model for DLAs includes most of the updates presented in \cite{Ho:2020}, but for computational reasons finds only one DLA per spectrum and does not include the sub-DLA model. Section~\ref{sec:training} summarizes our training set. Our results are presented in Section~\ref{sec:result} and we conclude in Section~\ref{sec:conclusion}.
Our redshift estimation code is available on github at \url{https://github.com/sbird/gp_qso_redshift}. Our DLA model with redshift estimation may be found at \url{https://github.com/sbird/gp_dla_detection/tree/zqsos2}.

\section{A Gaussian Process Model for QSOs}
\label{sec:gpmodel}

Quasar emission spectra are complex functions which do not have a known
closed parametric form. Our method builds a model for the expected shape of a quasar emission spectrum $f(\lambda)$. We use a Gaussian process \citep[GP:][]{williams2006gaussian}, a non-parametric framework able to model complex continuous functions. Importantly, a Gaussian process can describe how variations in the observed spectra are correlated as a function of wavelength. The learned model will naturally include information describing the presence of emission lines. The training set for our model is SDSS DR12 with Broad Absorption Line (BAL) quasars removed and $z_\mathrm{VI} < 2.15$. We use SDSS visual inspection redshift estimates during training. However, the trained model is applicable to larger, unlabelled, datasets. After training, the learned model is used to evaluate the likelihood function of each quasar spectrum as a function of redshift. Our point redshift estimate is located at the maximum a posteriori value of the likelihood function and the redshift uncertainty is given by $95\%$ confidence intervals.

A GP is a generalization of the Gaussian distribution which
describes random functions, rather than random vectors.
Naively, we can think of a GP as a Gaussian distribution extended over an
infinite number of dimensions.  It is described by a mean {\em function},
$\mu(\lambda)$ and a covariance {\em function}, $K(\lambda_1,\lambda_2)$.
The mean describes the average value of a draw (of a function) from the
GP.  The covariance describes the correlations between any two points on
the function, $f(\lambda_1)$ and $f(\lambda_2)$.
%
If a fixed set of regressors, for example $\lambda_1, \lambda_2,
\dots, \lambda_m$, is selected, the random function
evaluated at these values generates a set of (dependent) random variables:
$f(\lambda_1), f(\lambda_2), \dots, f(\lambda_m)$.  In a GP, these random
variables are jointly Gaussian.  Their means are just the application of
$\mu$ to the independent values, and the covariance matrix is similarly
constructed: $E\left[f(\lambda_i)\right] = \mu(\lambda_i)$ and
$\text{covar}\left[f(\lambda_i),f(\lambda_j)\right] = K(\lambda_i,\lambda_j)$.

There are no off-the-shelf Gaussian process covariance functions able to model the complex shape of a quasar. We thus learn a covariance function from the training data. Our model assumes that the emission spectrum from a QSO (in its rest
frame), $y$, is drawn (independently from $\zqso$) from a Gaussian process
with mean function $\mu$ and a covariance function $K$, which we denote as
\begin{align}
	p(y) &= {\cal N}(y; \mu, K)\,\,.
	\label{eq:basegp}
\end{align}

We choose to build our GP model at the rest-frame $\rszqso(\lambda_{OBS})$.
We, therefore, can capture the covariance between different emission lines from different quasars by
setting them onto the same rest-wavelength pixels.
The relationship between the rest-frame and observed-frame is
\begin{equation}
   \rszqso(\lambda_{OBS}) = \frac{1}{1 + \zqso} \lambda_{OBS}\,\,.
\end{equation}

The Gaussian process describing the QSO spectrum can
be transformed into the observed-frame, and remains a Gaussian process.
Letting $\rf{y}$ be the emission spectrum in the observed-frame,
\begin{align*}
	p(\rf{y}) &= {\cal N}\left(\rf{y}; \comp{\mu}{\redshift_\zqso},
		\comp{K}{\redshift_\zqso}\right) \\
		(\comp{\mu}{\redshift_\zqso})(\lambda_{OBS}) &= \mu(\redshift_\zqso(\lambda_{OBS})) \\
		(\comp{K}{\redshift_\zqso})(\lambda_1,\lambda_2)
		&= K\left(\redshift_\zqso(\lambda_1),\redshift_\zqso(\lambda_2)\right)\,\,.
\end{align*}

The observed spectrum, \obs, is equal to $\rf{y}$, but after absorption between the observer and the quasar and additive noise from the observational instrument.
Calculating the scale factor $\frac{1}{1 + \zqso}$ requires knowledge of the quasar redshift.
Let $\Data = (\lambda_{OBS}, \obsflux)$ be a set of quasar observations in the observed-frame,
where $\lambda_{OBS}$ is the set of wavelengths in the observed-frame, and $\obsflux$ is the set of observed flux.
We learn our GP model $\mzqso$ at $\rszqso(\lambda_{OBS})$ using a training set of observations with known quasar redshifts,
$\Data = (\rszqso(\lambda_{OBS}), \obsflux, \zqso)$, where $\zqso$ is the redshift estimated by the SDSS pipeline.
After we learn the GP model $\mzqso$, we use observations outside the training set
$\Data = (\lambda_{OBS}, \obsflux)$ to validate our $\mzqso$.


We assume that absorption between the observer and the quasar and
additive noise from the observational instrument are independent of each other
and that both are uncorrelated between wavelength bins.
The instrument noise is modeled using a
Gaussian process with a zero mean function and a ``diagonal'' covariance
kernel. $K(\lambda_1,\lambda_2)$ is zero if $\lambda_1$
and $\lambda_2$ are not equal (or almost equal). Instrument noise is a
property of the survey, and is not learned during training.
If $K_N$ is the kernel for the instrument noise, the observed spectrum, \obs,
is also drawn from a Gaussian distribution if we condition on $\zqso$:
\begin{align*}
	p(\obs |  \zqso) &= 
		 {\cal N}\left(\obs; \comp{\mu}{\redshift_\zqso},
		 (\comp{K}{\redshift_\zqso})\! +\! K_N\right)
    \label{eq:purezmodel}
\end{align*}

Section~\ref{sec:absorption} describes neutral hydrogen absorbers in the intergalactic medium, which are treated separately.
As they do not strongly affect the shape of the peaks which dominate the
redshift estimation, we neglect them except when finding DLAs. We have not attempted to model BAL and have removed BAL quasars from the sample.

\section{Learning A GP for Redshift-Estimation}
\label{sec:pure_z}
In this section, we describe the modelling decisions we made to extend
our Gaussian process model, $\mzqso$, for quasar redshift estimation.
$\mzqso$ is a lightweight GP model and may be sampled to obtain the likelihood of the quasar redshift, $\zqso$. The shape of this likelihood in turn produces $p(\zqso \mid \obsflux, \mzqso)$, the posterior distribution for $\zqso$.

The null model $\mzqso$ contains information describing the average shape of a quasar. A minimal modification of \cite{Garnett:2016} would fit this null model to different quasar redshifts. We found however, that this minimal modification does not have sufficient information to fit the quasar. We thus modify it in two important ways. First, we extend the modelled Gaussian process range to $910 - 3000$ \AA, in order to encompass more emission lines, especially MgII. Second, we augment the model to explicitly model the likelihood of observations outside the modelled redshift range. There is thus some likelihood component for all observations and so probabilities are comparable for the same spectrum across multiple redshifts.

\subsection{Redshift prior}
In this paper we treat $\zqso$ as a parameter to estimate, rather than a known value. We place a
bounded uniform prior on the parameter $\zqso$, $p(\zqso)$:
\begin{equation}
      p(\zqso) = \uniform[\zqsomin - z_\epsilon, \zqsomax + z_\epsilon],
   \label{eq:uniform_prior}
\end{equation}
where $\zqsomin$ and $\zqsomax$ are the minimum and maximum quasar redshifts. For our SDSS sample they are $2.15$ and $6.44$, respectively. We extend the prior range by a small amount ($z_\epsilon = 3000$ km/s) on either side to ensure that no samples
lie on the prior boundaries. We use a uniform prior rather than a data-driven prior to demonstrate that our method
is applicable to arbitrary quasar spectra within the prior range, rather than just the SDSS dataset.\footnote{Note that for the DLA finding problem we use a different, data-driven, prior as we integrate out $\zqso$ to find $\lognhi$ and $\zdla$.}


\subsection{Extended Model Range}
\label{sec:extendedmodel}

The original modelling range of $\mzqso$ ran from the rest-frame Lyman limit ($910$~\AA)~to the rest-frame {\Lya} ($1216$~\AA). We extend this range to cover much of the metal line region. In the rest-frame
\begin{equation}
   \rszqso(\lambda_{OBS}) = \lambdarest  \in [ 910 \AAtext, 3000 \AAtext ].
   \label{eq:lambda_spacing}
\end{equation}
An extension to $3000\,\AAtext$ allows us to include the MgII emission line ($2799 \AAtext$). MgII is a particularly valuable emission line as it is the least affected by systemic velocity shifts \citep{Hewett:2010, Shen:2016}.
The pixel spacing remains the same as that of \cite{Garnett:2016} with $\Delta \lambda = 0.25 \AAtext$, giving us $8\,361$ pixels in our GP mean vector.

Blueward of the Lyman limit, the occasional presence of
strong absorption from a Lyman limit system introduces substantial variance into the model, so that it has little redshift constraining power. Furthermore, this region is hard to train. Only relatively rare $\zqso > 3.7$ quasars contain rest-frame data at $z < 910$\AA. We thus exclude the region blueward of the Lyman limit from the modelling range of the Gaussian process.

To model the relationship between quasar flux measurements and the true QSO emission function, we have to include the correlation between emission lines $K$ and the instrumental noise $\Knoise$.
When we are only interested in estimating redshift, we do not include the
model for neutral hydrogen absorption (``{\Lya} absorption noise'' in our earlier papers). This model affects only the continuum blueward of the \Lya~peak, which has relatively
large instrumental noise compared to the metal-line region and is thus sub-dominant when estimating redshift. We have confirmed that this approximation does not significantly affect our results, yet it reduced the training time for the model by a factor of $\sim 20$.

\subsection{Observed Data Outside GP Range}
\label{sec:unmodelleddata}

As we do not model the entirety of the quasar spectrum, our likelihood is incomplete. We would like to evaluate the marginal likelihood of the GP to estimate $\zqso$. However, to ensure that we can compare posterior probabilities at different redshifts, we need to provide a likelihood function for the data not modelled by the main GP. Otherwise, as different observations fall into the model, likelihoods are evaluated on different subsets of the data. To avoid this problem, we implemented an explicit model for observed data outside the Gaussian process model boundaries. All observed data is thus accounted for in the extended model.

To illustrate the need for this model, consider when emission peaks are redshifted out of the GP model range. A $z \sim 2.5$ quasar assumed to be at $z = 5$ will have the emission corresponding to the \Lya~emission peak at $1216$~\AA~incorrectly appear at $700$~\AA, outside the modelled rest-frame. As the peak is now outside the rest-frame, $\mzqso$ applies no penalty
for not predicting the emission peak and may incorrectly prefer a high
redshift.

Our explicit extra model assumes that the emission spectrum in the rest-frame bluewards of {$910$~\AA} are drawn independently and identically from a Gaussian distribution with a constant variance. We make the same assumption for those emission spectrum values redwards of the GP model's range. These ``out-of-GP'' emission fluxes are subject to the same instrument noise and absorption as the rest of the spectrum, after being transformed to the observer frame. However, they have no correlations with each other or with the flux modelled by the GP in $910-3000$~\AA.


\newcommand\mured{\mu_{\text{red}}}
\newcommand\mublue{\mu_{\text{blue}}}
\newcommand\stdred{\sigma_{\text{red}}}
\newcommand\stdblue{\sigma_{\text{blue}}}
\newcommand\redset{{\cal X}_{\text{red}}(\zqso)}
\newcommand\blueset{{\cal X}_{\text{blue}}(\zqso)}
\newcommand\inststd[1]{\sigma_{#1}}

The mean and standard deviations
of these two Gaussian distributions are optimized
for during training. We define $\mured$ and $\stdred$ to be the
mean and standard deviations of the ``out-of-GP'' model for the redward end.
If $\stdred$ is known, the maximum likelihood estimate for $\mured$ can be
computed in closed form:
\begin{align}
	\mured &= \frac{\sum_i \rho_i x_i}{\sum_i \rho_i}\,,
\intertext{where}
	\rho_i &= \frac{1}{\inststd{i}^2 + \stdred^2}\,.
\end{align}
Here $i$ ranges over observations in the training set that
fall redwards of the Gaussian process model and $x_i$ is the observed flux
(recall that the training data have known $\zqso$ values). $\inststd{i}$
denotes the standard deviation of the instrumental noise for observation
$i$. Thus each observation, $x_i$, is drawn independently from a normal
distribution with mean $\mured$ and variance $\stdred^2 + \inststd{i}^2$.

To find $\stdred$, we conduct a line search to find the maximum likelihood, using
the above substitution for $\mured$ in terms of $\stdred$ in the likelihood.
The resulting function (ignoring constants) to be optimized is
\begin{align}
	\log \mathcal{L} &= \sum_i
	\rho_i\left(x_i - \mured\right)^2\,
	-\log \rho_i
\end{align}
where $\rho_i$ and $\mured$ both depend on $\stdred$, the quantity to be tuned.
Empirically, this likelihood is concave and easy to maximize.
The fitting procedure for the blueward end model
is identical, but on a different set of fluxes.



\subsection{Quasar Normalisation}
\label{sec:normalization}

The observed magnitude of a quasar depends on its luminosity distance and the properties of the black hole.
To allow a single GP model to describe the observed flux $\obsflux$, we normalize the flux measurements.
\cite{Garnett:2016} chose to normalize at an absorption free region between $1310$ {\AA} and $1325$ {\AA} in the rest-frame.
Here we change the normalization range to $1176$ {\AA} $\sim$ $1256$ {\AA} for building $\mzqso$,
normalizing all spectra at the same \Lya~peak amplitude.

We choose to normalize the amplitude of the quasar spectrum to the {\Lya} peak region, $1216 \pm 40 \AAtext$. We found empirically that this produced the most accurate quasar redshift estimation during our validation experiments. The position of the {\Lya} peak is highly variable, which may at first make it seem a poor choice for normalization. We emphasise however that we only use the peak height, and not the peak position, to normalize the overall quasar continuum flux\footnote{Interestingly, the automated quasar continuum estimator of \cite{Reiman:2020} also normalizes continua using the height of the \Lya~peak}. The variability of the line is encoded in the GP covariance function, see Figure~\ref{fig:gpcovar}. We speculate that normalising to {\Lya} performs well because the strength of the {\Lya} line minimizes the impact of instrumental noise in the normalizing region on the continuum normalization. As the {\Lya} line is broad the normalization is also reasonably stable to small changes in $\zqso$.

We tried normalizing the quasar to the median continuum and to the CIV peak. Normalizing to the continuum led to complex unphysical structure in the learned covariance matrix and poor results. Normalizing to the CIV peak gave a tolerable covariance, but produced about a factor of two more redshift estimation failures than normalizing to the \Lya~peak.

During the testing phase, the observed flux $\obsflux$ has to be normalized for each redshift possibility, as
the region of observed spectrum which corresponds to the normalization region in the rest-frame changes with assumed quasar redshift.
We transform the spectrum as follows:
\begin{equation}
   \begin{split}
      \obsflux  &\leftarrow \obsflux / \xnorm(\zqso)\\
      \xnorm(\zqso) &=  \mathrm{median} \left[ \yobs(\rszqso(\lambda_{OBS}) \in [1176 \AAtext, 1256 \AAtext]) \right].
   \end{split}
\end{equation}
This transformation is done separately for every redshift sample, $\zqso$. Thus the normalization is redshift dependent and the likelihood depends only on the normalized flux.
$\mzqso$ is again defined on the rest-frame wavelengths $\rszqso(\lambda)$ and the normalized flux $\emissiony$, which is the emission spectrum without any intervening DLAs.

\begin{figure*}
\centering
   \includegraphics[width=2.2\columnwidth]{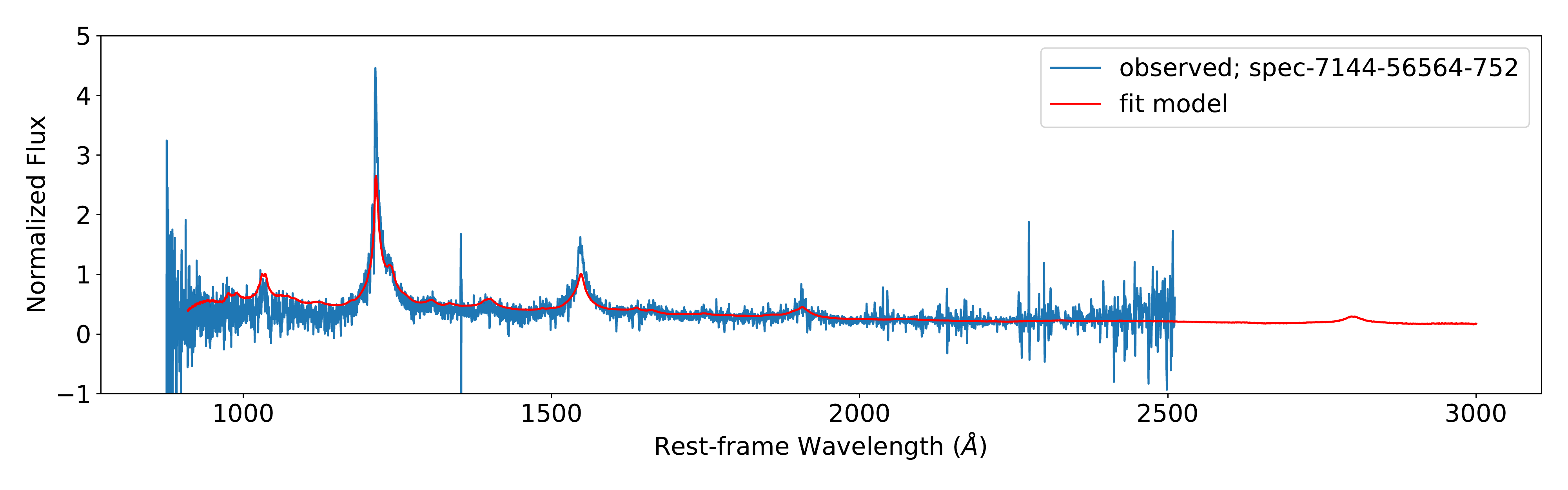} \\
   \includegraphics[width=2.2\columnwidth]{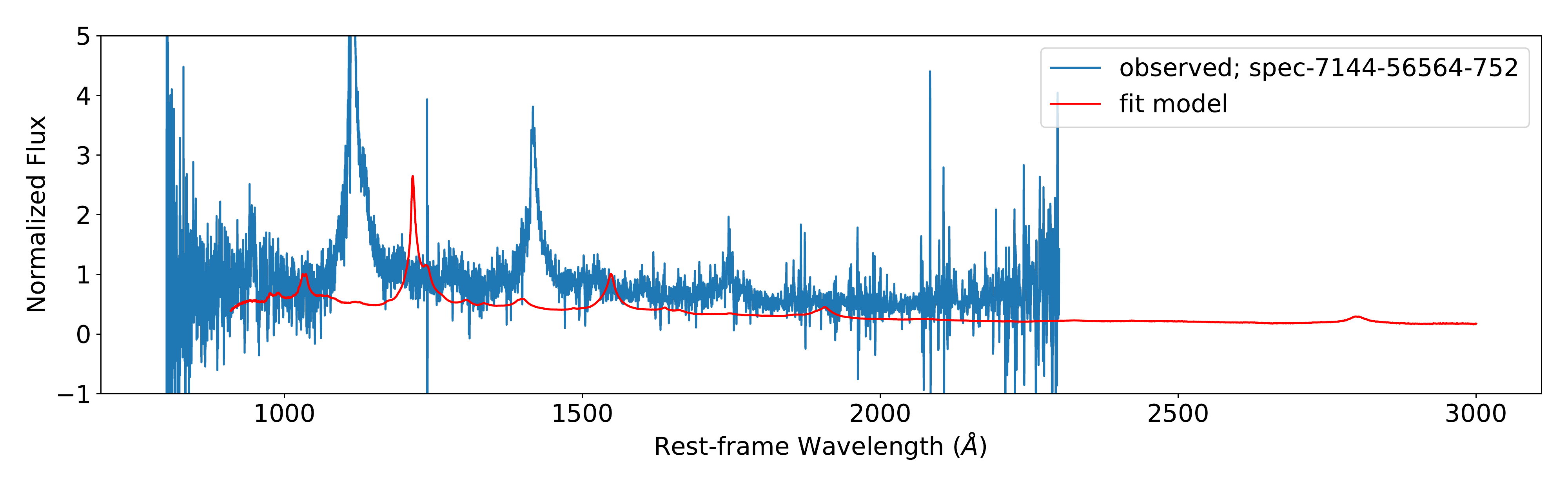}
   \caption{An example spectrum for our redshift estimation model.
   \textbf{Red curve:} the GP model mean for redshift estimation.
   \textbf{Blue curve:} the raw observed flux, after normalization at the range of $1216 \pm 40${\AA}.
   \textbf{(Top)} At the maximum likelihood redshift of this quasar. \textbf{(Bottom)} At an incorrect quasar redshift.
   Note that the normalization of the quasar is incorrect. Normalizing in the $1216 \pm 40 \AAtext$ region
   can introduce an additional penalty for incorrect redshifts.
   }
   \label{fig:this_mu_example}
\end{figure*}

An incorrect normalisation factor, $\xnorm$, substantially changes the likelihood of the quasar. Thus in most cases, the normalization factor
is close to the true $\xnorm(\zqso_\textrm{true})$
if and only if the $\zqso = \zqso_\textrm{true}$, inducing an
additional penalty in a $\zqso$ sample which is not close
to the true quasar redshift. However,
the roughly flat shape of the average quasar continuum means that
fitting different emission peaks to \Lya~still produces a plausible normalization.
Figure~\ref{fig:this_mu_example} illustrates such an incorrect normalization
from choosing a wrong $\zqso$.

\subsection{Redshift Estimation Model Summary}
\label{sec:zestimatesummary}

Combining all modelling decisions, the model prior for an observed QSO emission is
\begin{equation}
   \begin{split}
	   p&(\emissiony \!=\! \obsflux\mid  \mzqso, \zqso) \\
	 &= \normal\left(\frac{\obsflux}{\xnorm(\zqso)}; \murszqso, \Krszqso +
	   \frac{\Knoise}{\xnorm(\zqso)^2} \right) \\
	   &\phantom{==}\times
	   \prod_{\mathclap{\lambda\in \redset}}
	   \normal\left(\frac{x(\lambda_{OBS})}{\xnorm(\zqso)};\mured,\stdred^2+\frac{\inststd{\lambda}^2}{\xnorm(\zqso)^2}\right) \\
	   &\phantom{==}\times
	   \prod_{\mathclap{\lambda\in \blueset}}
	   \normal\left(\frac{x(\lambda_{OBS})}{\xnorm(\zqso)};\mublue,\stdblue^2+\frac{\inststd{\lambda}^2}{\xnorm(\zqso)^2}\right)
   \end{split}
   \label{eq:model_evidence_null}
\end{equation}
where $\redset$ are the set of observed wavelengths which fall outside of
the Gaussian process model when transformed into a rest-frame of $\zqso$.
By sampling from the parameter prior $p(\zqso)$, this model prior serves
as a likelihood function for a QSO observation being at a given $\zqso$.

The first $\normal$ is the density of a Gaussian process, evaluated on the
observations that fall within the Gaussian process model.  The last two $\normal$
are standard normal densities on the scalar values of the observations
that fall outside the Gaussian process model.
The observed instrumental noise is normalized by $\xnorm(\zqso)^2$, so that
Eq.~\ref{eq:model_evidence_null} shows the noise kernel $\Knoise$ after
normalization. $(\murszqso, \Krszqso)$ denotes the mean function and
covariance kernel in the quasar rest-frame.  The mean function and
covariance function are only modelled within the range based on
Eq.~\ref{eq:lambda_spacing}.  At the testing phase, we thus only evaluate
the GP likelihood of $\obsflux(\lambda)$ inside the modelling window.
We use the quasi-random Halton sequence to generate
$10^4$ samples of $\zqso$ from our prior for $p(\zqso)$.

\subsection{Learning the flux mean vector and covariance}
\label{subsec:learn_covariance}

\begin{figure*}
\centering
   \includegraphics[width=2\columnwidth]{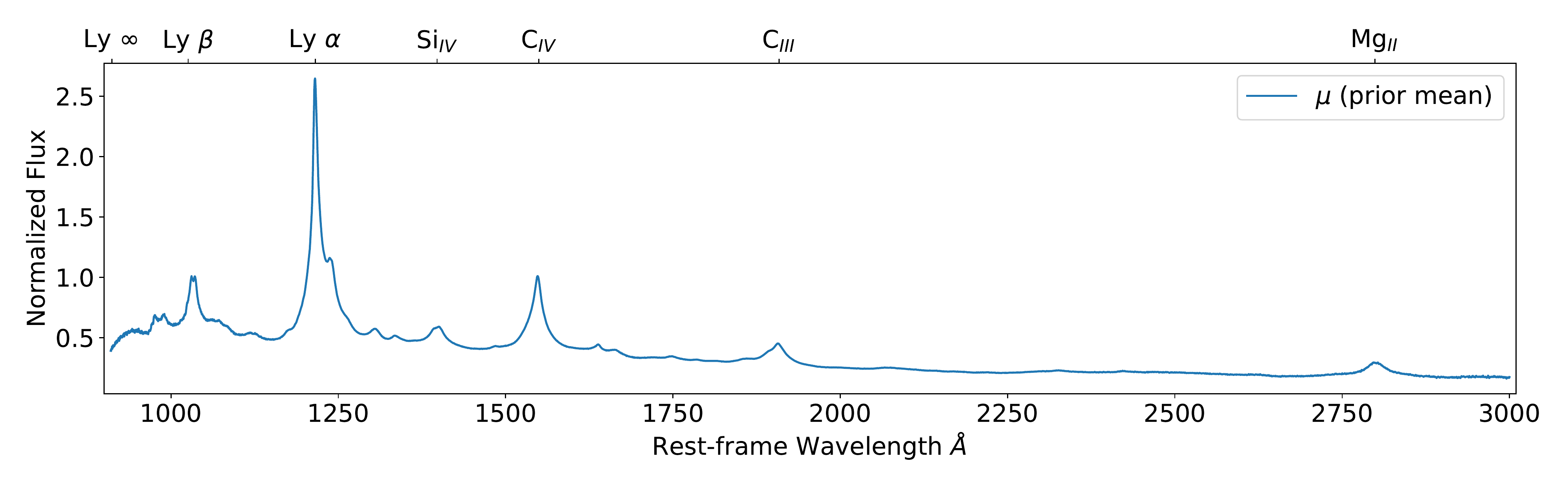}
   \caption{The estimated mean vector for our rest-frame quasar model, found by taking the mean value for each interpolated value across all rest-frame spectra in the training set. The rest-frame locations of common emission lines are shown in the upper axis.}
   \label{fig:gpmuvec}
\end{figure*}

\begin{figure}
   \includegraphics[width=\columnwidth]{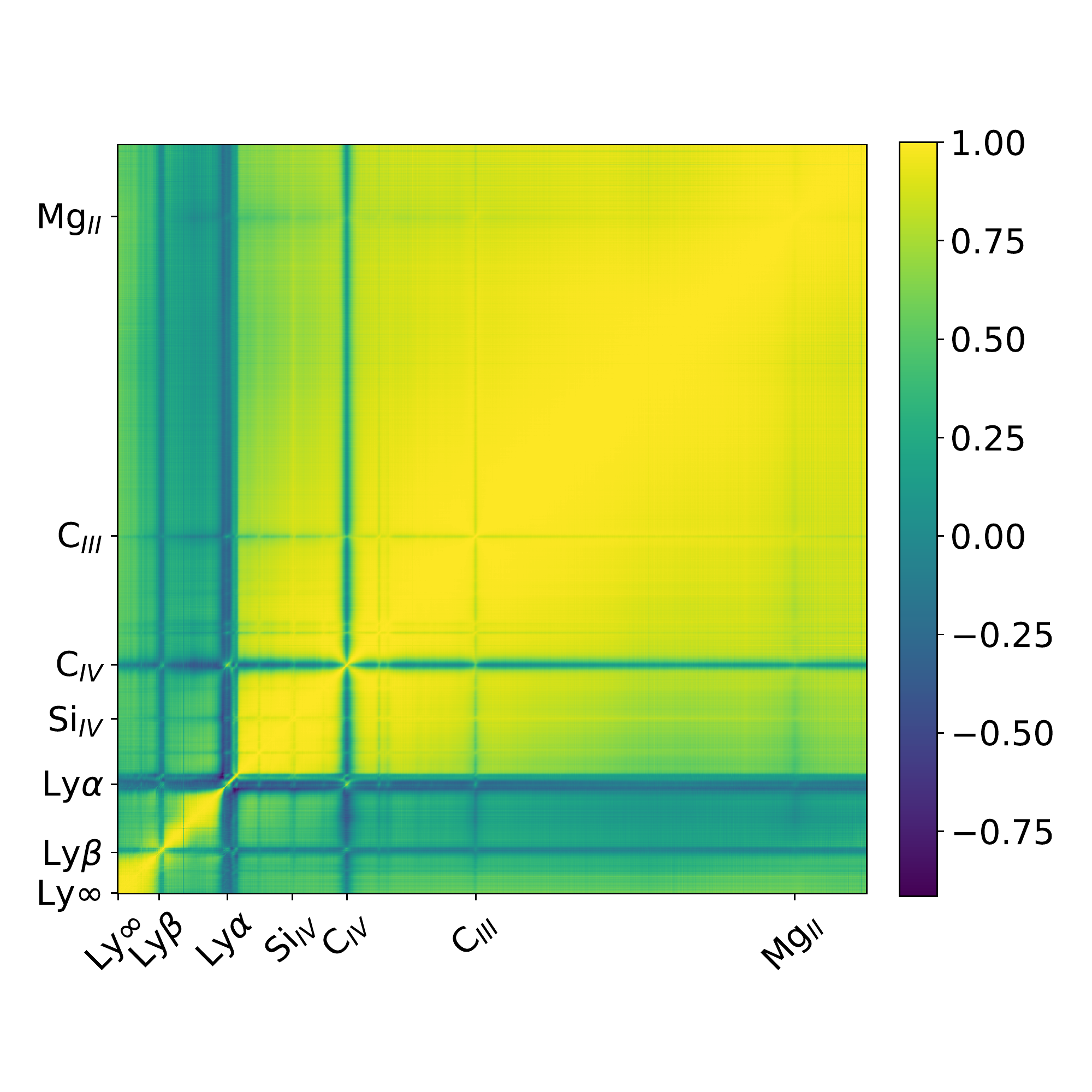}
   \caption{The trained correlation matrix $\Kvec$, with the $\lambda$ range from $910$\,{\AA} to $3\,000$\,{\AA}.
   We have normalized the diagonal elements to be unity. The values in the matrix range from $-1$ to $1$, representing the correlation between $\lambda$ and $\lambda'$ in the {\qso} emission function.}
   \label{fig:gpcovar}
\end{figure}

In this Section, we describe how we learn $\mu$ and $K$ of our GP model $\mzqso$.
Both are discretized.  That is, we model $\mu$ as a piecewise-constant function whose ``pieces'' are of
fixed widths.  Thus, its parameterization is as a vector of the mean values over each piece.  $K$ is similarly
discretized as a matrix.

Each observed spectrum is transformed to the rest-frame and the values interpolated to the mid-points of the
piecewise constant representation.  Each element of the $\mu$ vector is estimated as the mean of all available\footnote{Some
observations are missing or have instrumental noise variance larger than $4^2$ and are omitted.}
rest-frame flux
values at the same wavelength.
The learned mean from the data is shown in Figure~\ref{fig:gpmuvec}, and clearly shows the expected series of metal emission lines.

To acquire the kernel matrix $K$, 
we assume the same likelihood as \cite{Garnett:2016} except (for now) excluding the absorption noise:
\begin{equation}
   \begin{split}
      p(\Yvec &\mid \mzqso, \zqso)
      = \prod_{i=1}^{N_{\textrm{spec}}}
      \mathcal{N} (\emissiony_i; \murszqso, \Krszqso + \Knoise ),
   \end{split}
   \label{eq:likelihood_whole_dataset}
\end{equation}
where $\Yvec$ represents the matrix of all observed flux measurements in the training set, each transformed into the rest-frame on a
standard grid.
The covariance matrix $K$ is learned via the low-rank decomposition
\begin{equation}
   \Kvec = \Mvec \Mvec^{\top}.
   \label{eq:low_rank_decomposition}
\end{equation}
$\Kvec$ is the kernel $(\Krszqso)$, conditioned on the rest-frame wavelength pixels we defined before, and
$\Mvec$ is an ($N_{\textrm{pixels}} \times k$) matrix, with $N_{\textrm{pixels}} = 8\,361$ and $k = 20$.

Our kernel is trained by optimizing the values of $\Mvec$ to maximize the likelihood given in Eq.~\ref{eq:likelihood_whole_dataset}.
We use the first $k$ principal components of $(\Yvec - \mu)$ as initial conditions.
With the much larger model range (and thus matrices) trained in this paper, the MATLAB PCA function often failed to find principal components. This was due to substantial missing or noisy data at the red side of the training set. To allow the PCA to converge, we replaced all such data, represented in our dataset by NaN, with the median value of the whole spectrum before taking the PCA. Although this kind of missing data imputation generally biases a PCA, in this case we are only using it as a starting point for our algorithm, and subsequently optimizing it away.
Optimization is still done using the unmodified $\Yvec$ and uses the same unconstrained optimization as in our earlier papers, except without gradients of the absorption noise model. 

Figure~\ref{fig:gpcovar} shows the learned kernel. The bottom left resembles the similar figure of \cite{Garnett:2016}, which was evaluated only in that range. The dark vertical lines in Figure~\ref{fig:gpcovar} show pixel areas which have strong correlations only in a narrow wavelength range. These areas drive the final redshift estimate and correspond to the locations of well-known major emission lines. Particularly strong are CIV (1549 \AA), Lyman-$\alpha$ (1216 \AA) and OVI (1034 \AA). Weaker signals are shown for MgII (2799 \AA), CIII (1909 \AA), SiIV (1397 \AA) and CII (1335 \AA). Although MgII is a famously reliable line \citep{Hewett:2010}, its presence in the correlation matrix is reduced because the emission line is low amplitude compared to the instrumental noise at long wavelengths. On the other hand, CIV, Lyman-$\alpha$ and OVI are extremely strong emission lines and thus more visible. The width and variability in the {\Lya} line position shows up as the width of the correlation band around $1216$~\AA. The similar width of the CIV line may be due to the code learning the correlation between CIV equivalent width and line blueshift \citep{Gaskell:1982, Sulentic:2007, Richards:2011, Mason:2017}.

\section{DLA Finding Model}
\label{sec:dlafind}
%

In this Section, we describe how our quasar redshift estimator can be extended to find DLAs,
while marginalizing out quasar redshift uncertainty. We take the model
presented in Section~\ref{sec:pure_z} and combine it with the DLA model from \cite{Ho:2020}.
The most important changes to the model are the inclusion of a model for Lyman-series
absorbers along the line of sight to the quasar (Section~\ref{sec:absorption}) and
an explicit model for DLAs (Section~\ref{sec:dlamodel}).

We do not use the uniform prior quasar redshift distribution from Section~\ref{sec:pure_z}.
Instead we use as a prior a $150$ bin histogram of $\zqso$ from the training data.
We have checked explicitly that pure redshift estimation with this prior leads to similar results
as the uniform prior, with some minor sampling artifacts at high redshift.

\subsection{Lyman-series Absorption}
\label{sec:absorption}

Following \cite{Ho:2020}, we supplement our instrumental noise model with an additional variance term
to account for absorption from Lyman series lines, especially the \Lya~forest. We model Lyman series absorption
as Gaussian noise with a redshift dependent mean and variance, but no inter-pixel correlations. Our Gaussian process model for
redshift estimation from Section~\ref{sec:pure_z} is thus modified by adding the diagonal absorption noise kernel $K_A$:
\begin{align*}
	p(\obs | \zqso) &= 
		 {\cal N}\left(\obs; \comp{\mu}{\redshift_\zqso},
		 (\comp{K}{\redshift_\zqso})\! +\! K_A + K_N\right)
\end{align*}
As \Lya~forest absorption is only possible in the region of the spectrum bluewards of the \Lya~line in the quasar rest-frame, we include an indicator function in $K_A$, so that absorption is zero for $\lambda_{REST} > 1216$~\AA.

Evolution of the \Lya~forest flux with redshift is included by assuming the absorption noise has a power-law redshift dependence, so that $K_A$ is given by:
\begin{align*}
	K_A(\lambda_{REST},\lambda_{REST}') &= \delta(\lambda_{REST}-\lambda_{REST}')  \\
	\times I(\lambda_{REST} < 1216 &)
	\left(1 - \exp\left(-\tau_0 (1 + z_\mathrm{Lya})^\beta\right) + c_0\right)^2,
\intertext{where}
z_\mathrm{Lya} &= \frac{\lambda_{REST}}{1216}\left(1+\zqso\right) - 1\,\,.
\end{align*}
$c_0$, $\tau_0$, and $\beta$
are constants, and $z_\mathrm{Lya}$ is the redshift of Lyman-$\alpha$ at the
observed wavelength. Hence our model depends on the redshift of the
quasar as well as the redshift of Lyman-$\alpha$ along the line of
sight.

One unphysical feature of our absorption noise model is that, because
Gaussian noise is symmetric, it assumes emission is as likely as absorption. This is
particularly dangerous at high redshift, where the average absorption in a quasar spectrum is substantial.
As we showed \citep{Ho:2020}, we can account for this by modifying
the quasar mean vector to match the observed mean flux of the \Lya~forest.
We assume an effective optical depth $\tau_0 (1 + z_\mathrm{Lya})^\gamma$ following \cite{Kim07}:
\begin{align}
      a(z_\mathrm{Lya}) &= \exp{ ( -\tau_0 ( 1 + z_\mathrm{Lya} )^{\gamma} ) } \\
          &= 0.0023 \times \exp{(1 + z_\mathrm{Lya})^{3.65}},
   \label{eq:effective_optical_depth}
\end{align}
We include absorption for the first $6$ Lyman series lines, accounting for the different absorption coefficients. We account for the mean suppression from Lyman series absorption in our redshift-dependent noise model $K_A$. The complete {\gp} model mean, written as a function of observed-frame wavelength $\lambdaobs$, for each spectrum is thus:
\begin{equation}
   \begin{split}
      a(\lambdaobs/\lambda_{\textrm{Ly}\alpha}-1) \times (\mu \circ \rszqso) (\lambdaobs)
   \end{split}
   \label{eq:mean_flux_kim_prior}
\end{equation}

The parameters for the redshift-dependent component of
the absorption noise vector were
\begin{align}
   c_0 = 0.3050; \tau_0 = 1.6400 \times 10^{-4}; \beta = 5.2714.
   \label{eq:learned_absorption_params}
\end{align}
Once the absorption model is included, there are degeneracies between different hyperparameters of the GP kernel. This increases training time and means that the training does not technically converge. Instead it moves along a trough with the maximum likelihood changing by less than $0.1\%$. Our trained model stopped training after $1500$ minimization steps, although early iterations were trained to $3000$ iterations with little difference in the kernel function.

\subsection{DLA model}
\label{sec:dlamodel}

We introduce an alternate model for DLA spectra following \cite{Garnett:2016}. Either the DLA or no-DLA model is chosen by Bayesian model selection. The presence of a DLA is indicated by its Voigt
profile, which includes absorption due to higher order Lyman lines:
\begin{align*}
	\rfdla{y}(\lambda_{OBS}) &= \rf{y}(\lambda_{OBS}) \exp\left(-\tau(\redshift_{\zdla}(\lambda_{OBS}); \colden)\right)\,\,.
\end{align*}
Here $\rfdla{y}$ is the emission spectrum after DLA absorption and $\tau(\lambda; \colden)$ is the Voigt profile
for column density $\colden$ at wavelength $\lambda$.
The DLA model (\DLA) has two parameters: the DLA redshift $\zdla$ and the DLA
column density $N_\mathrm{HI}$.
We take the prior redshift distribution of the DLA, $p(\zdla\mid\DLA,\zqso)$,
to be uniform between a region $3,000$ km/s redwards of the Lyman limit at $910$~\AA~and $3,000$ km/s bluewards of $\zqso$.

The prior distribution over the column
density, $p(\colden\mid \DLA)$, is modelled as a log-normal distribution. We use a kernel density estimate from the DR 9 sample, mixed with a
uniform distribution (Eq.~51 of \cite{Garnett:2016}). We do not include the sub-DLA model of \cite{Ho:2020}.

\subsection{Model Inference}
\label{subsec:modinf}

Our full model is
\begin{align*}
	p(x,\zqso,\noDLA) &= p(\zqso)\times \Pr(\noDLA\mid \zqso) \nonumber \\
		&\times p(x\mid \noDLA,\zqso) \\
	p(x,\zqso,\DLA,\zdla,\colden) &=
		p(\zqso)\times \Pr(\DLA\mid \zqso) \nonumber \\
		&\hspace{-4em}\times p(\zdla\mid \zqso,\DLA) \times p(\colden\mid \DLA) \nonumber \\
		&\hspace{-4em}\times p(x\mid\DLA,\zqso,\zdla,\colden)\,\,.
\end{align*}
$\zdla$ and $\colden$ can be marginalized out to obtain
\begin{align*}
	p(x,\zqso,\DLA)\! &=\! \iint p(x,\zqso,\DLA,\zdla,\colden)\,\,d\zdla\,d\colden\,.
\end{align*}
We are particularly interested in $\Pr(\DLA\mid x)$, the probability of a DLA given the observed
spectrum, and $p(\zqso\mid x)$, the distribution of the quasar redshift given the observed
spectrum. We calculate these conditional marginal distributions as follows.
\begin{align}
	p(\DLA,x) &= \int p(x,\zqso,\DLA)\,\,d\zqso \\
	p(\noDLA,x) &= \int p(x,\zqso,\noDLA)\,\,d\zqso \\
	\Pr(\DLA\mid x) &= \frac{p(\DLA,x)}{p(\DLA,x)+p(\noDLA,x)}
\intertext{and}
	p(\zqso\mid x) &\propto p(x,\zqso, \DLA) + p(x,\zqso, \noDLA)
	\label{eq:marg}
\end{align}
where the constant of proportionality in the last line makes $p(\zqso\mid x)$ integrate to $1$ over
$\zqso$.

Estimating the probability of a DLA requires a three-dimensional integral
over $\{\zqso, \zdla, \colden\}$ for $p(\DLA,x)$ and a one-dimensional integral over $\{\zqso\}$ for $p(\noDLA,x)$. As in Section~\ref{sec:pure_z}, we use the quasi-random Halton sequence
to generate 1- or 3-dimensional points as samples over the unit cube.
However, reflecting the higher dimensionality of our parameter space we draw $10^5$
samples per quasar instead of $10^4$.
We then transform them by the relevant inverse cumulatives to generate samples
from $p(\zqso)$ or $p(\zqso,\zdla,\colden)$ from which the integrals can be
numerically approximated (as the integrals can be transformed into
expectations with respect to these sampling distributions).
In this way, the likelihood of a DLA can be estimated without knowledge
of $\zqso$.

\subsection{Model Parameterization and Priors}
\label{subsec:modpri}

The full model requires the specifications of the following components. In the quasar rest-frame:
\begin{compactitem}
\item $\mu$: the mean quasar emission spectrum, and
\item $K$: the kernel of the Gaussian process for the emission.
\end{compactitem}
In the redshifted observer frame:
\begin{compactitem}
\item $K_A$: the diagonal non-DLA absorption variance, and
\item $K_N$: the diagonal instrument noise variance.
\end{compactitem}
Priors are given for
\begin{compactitem}
\item $p(\zqso)$: the redshift of a quasar,
\item $p(\colden\mid\DLA)$: the column density of the DLA, and
\item $p(\zdla\mid\DLA,\zqso)$: the DLA redshift distribution.
\end{compactitem}

\section{Training and Validation Data}
\label{sec:training}

The training set to learn our GP model $\mzqso$ for $\zqso$ estimate consists of the spectra observed by SDSS DR9. For DLA finding we also removed DLAs labelled in \cite{lee2013boss}. The validation data consisted of SDSS DR12, comprising $297,301$ quasar spectra. The following spectra were removed from both the training and validation set:
\begin{itemize}
   \item \texttt{$z_\textrm{VI} < 2.15$}: quasars with redshifts lower than $2.15$.
   \item \texttt{BAL}: quasars where SDSS found broad absorption lines.
   \item Spectra with less than 400 detected pixels.
   \item \texttt{ZWARNING}: spectra whose analysis by the SDSS pipeline flagged warnings.
   These spectra are usually not quasars, but represent some instrumental problem.
   We kept extremely noisy spectra with the \texttt{TOO\_MANY\_OUTLIERS} flag.
\end{itemize}
After these cuts, the remaining sightline catalogue is $158,979$ quasars. Given that the purpose of this paper is redshift estimation, it may seem circular to filter quasars with $z < 2.15$ from testing. However, these quasars do not contain DLAs, nor are they useful for \Lya~BAO. We examine these spectra further in Section~\ref{subsec:low_z} and show that our trained model still works reasonably well as long as the \Lya~emission peak (which we use for normalization) is inside the observed band, that is for $z \geq 1.9$.

Training the model requires a redshift estimate for the training data. Here we use the SDSS visual inspection redshift as it is available for the highest quasar fraction in the sample \citep{paris2018sloan}. Note, however, that visual inspection redshifts are not required. The model merely requires some redshift estimate. Future iterations could be trained using, for example, the DR12 redshift outputs of this paper, making the model fully self-hosting.


\section{Results}
\label{sec:result}

In this Section we describe the results of our algorithm run on the SDSS DR12Q dataset.
Section~\ref{sec:zresult} describes the results when estimating only quasar redshift.
Section~\ref{sec:dlaresult} also describes the results of our DLA finding.

\subsection{Redshift Estimation}
\label{sec:zresult}

In this section, we apply our QSO redshift model $\mzqso$ to SDSS DR12. We validate our ability to predict quasar redshift, $\zqso$. Although our model is fully Bayesian, we need a point estimate to compare to the SDSS catalogue redshift.
We use the \textit{maximum a posteriori} (MAP) of the sample posterior $p(\zqso \mid \obsflux, \mzqso)$, which is
equivalent to the maximum likelihood estimate (MLE) because we use a uniform prior for $p(\zqso)$.
We thus report the $\zqso$ sample with the highest likelihood
\begin{equation}
   \begin{split}
      \zmap = \mathrm{arg\,max}_{\zqso{_i}}
      &p(\emissiony(\zqso{_i})
      \mid \mzqso, \zqso{_i}),
   \end{split}
\end{equation}
where $\zqso{_i}$ is the $i^\textrm{th}$ Halton sequence sample.
The instrumental noise variance depends on $\zqso{_i}$ via normalization.

\begin{figure}
\centering
   \includegraphics[width=\columnwidth]{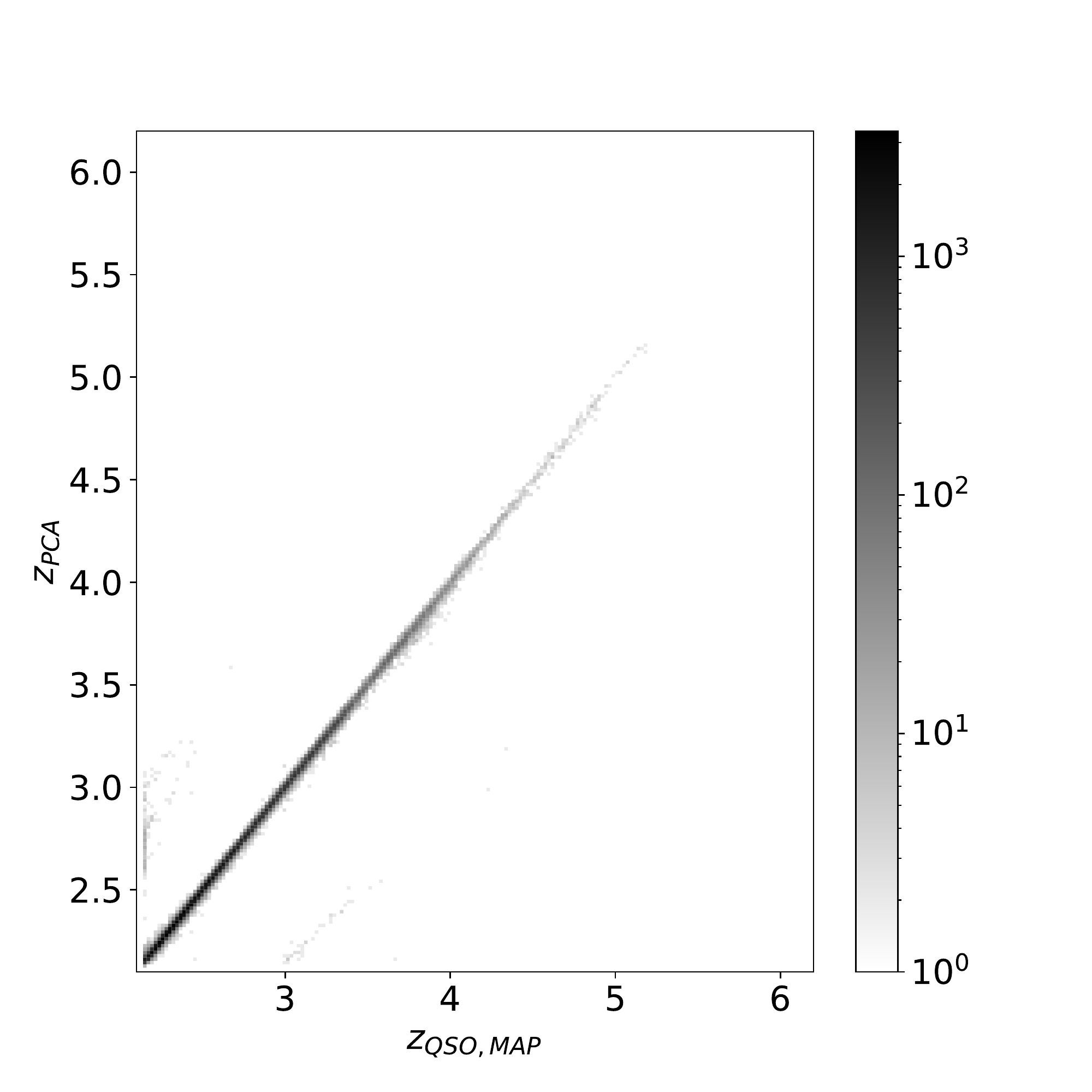}
   \caption{The MAP prediction of our catalogue, $\zqso$, versus the PCA redshift $z_{PCA}$ from the SDSS catalogue. The grey scale bar shows the number of quasars in each bin, using a logarithmic scale. The diagonal line in the middle of the plot shows a correct redshift estimation. Other diagonal lines correspond to occasional line fitting mistakes of our code.}
   \label{fig:z_map_vs_z_true}
\end{figure}

\begin{figure}
\centering
   \includegraphics[width=\columnwidth]{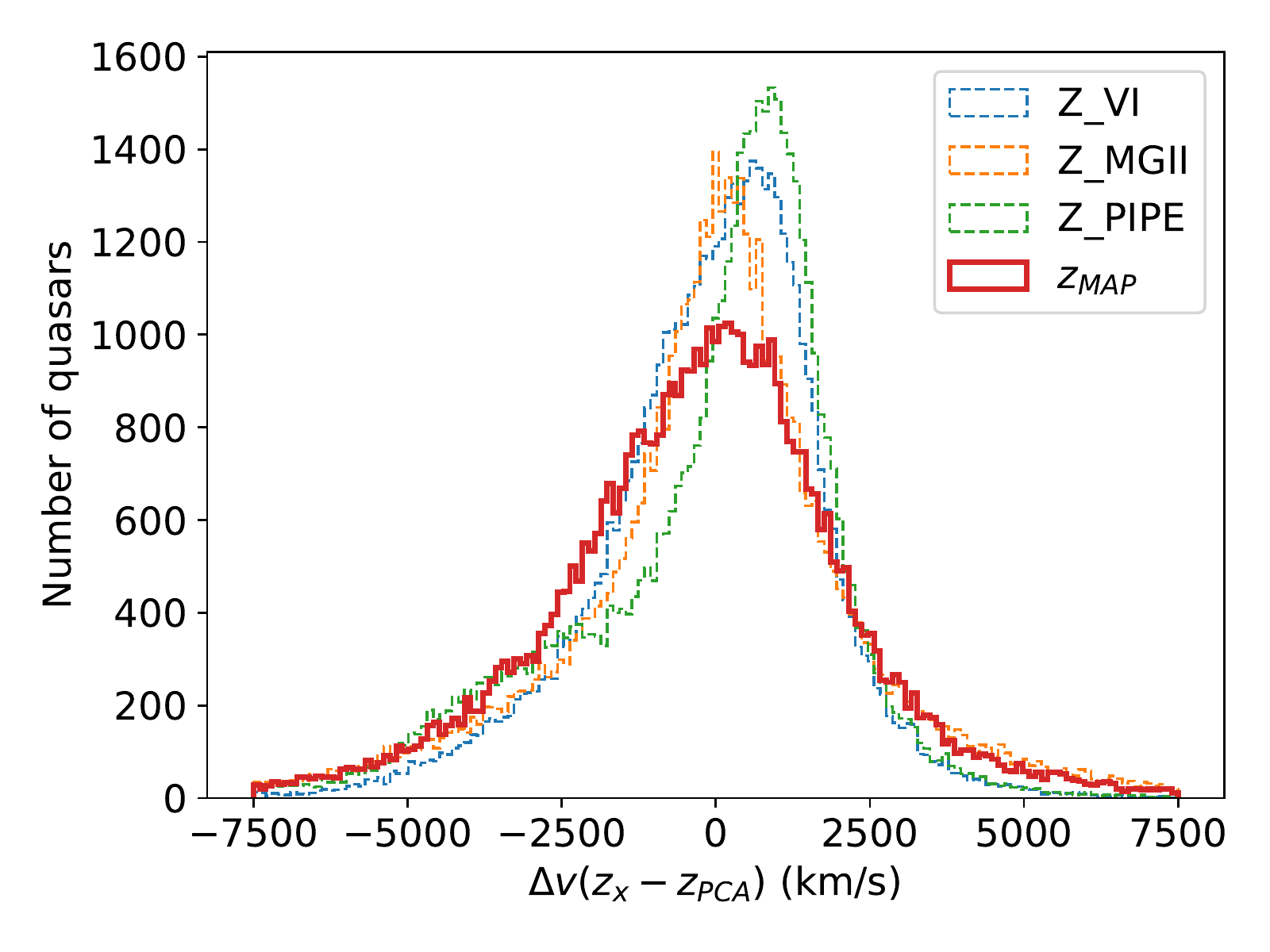}
   \caption{A histogram showing differences between the MAP prediction of $\zqso$ from our catalogue ($z_{MAP}$) and different $\zqso$ estimation techniques present in the SDSS catalogue. All methods are compared to the PCA based redshift, $z_\mathrm{PCA}$. We show results for only the $49776$ quasars with redshift estimates from all SDSS methods.}
   \label{fig:zmapvszpca}
\end{figure}

In Figure~\ref{fig:z_map_vs_z_true} we compare the MAP estimate of our catalogue, $\zmap$, to the reported PCA redshift $z_{PCA}$ in SDSS DR12. The two are generally in good agreement, as shown by the large number of quasars on the plot diagonal. There are a small number of cases where our model fits \Lya~using another emission peak, visible as the secondary lines above and below the main diagonal (note that Figure~\ref{fig:z_map_vs_z_true} uses a logarithmic scale). The above-diagonal line corresponds to \Lya~peaks being fit by OVI emission. This line is broad because OVI is in the \Lya~forest and so has large variance in our model. The below-diagonal line, which is narrower, corresponds to \Lya~peaks fit with CIV emission. There are also a few objects, of a density too low to be visible on the plot, where the code fits the OVI emission line to CIV. The rate at which our redshift estimation fails is low. Comparing to the PCA redshift we find that  $|z_{PCA} - \zmap| > 0.5$ for $0.38\%$, which is 603 out of 158560 quasar spectra. Comparing to the visual inspection redshift $z_{VI}$ gives similar results: $|z_{VI} - \zmap| > 0.5$ for 645 of 158979 spectra. For the more stringent bound of $|z_{VI} - \zmap| > 0.05$, the misfit rate rises to $0.99\%$. Other redshift measurements performed similarly, with $z_{CIV}$ having the lowest misfit rate ($0.35\%$) and $z_{PIPE}$ the highest ($0.43\%$).

Figure~\ref{fig:zmapvszpca} compares to other redshift estimation methods used in SDSS, following Figure~7 of \cite{paris2017sloan}. We show results only for the $49776$ quasars with redshift estimates from all SDSS methods.
Overall our technique performs similarly to the others. It is complementary in that it prefers lower redshifts than the PCA model $z_{PCA}$, while other methods prefer a generally higher redshift.

Our method has a median difference in redshift with $z_{PCA}$ of $-117$ km/s. The equivalent median differences between $z_{PCA}$ and other methods are $z_{VI}$ : 128 km/s, $z_{MgII}$ : 73 km/s, $z_{PIPE}$ : 380 km/s. Our technique is thus competitive in this metric. The standard deviation of this dispersion with $z_{PCA}$ is 17,000 km/s. The other methods score substantially better: $z_{VI}$ : 1800 km/s, $z_{MgII}$ : 2500 km/s, $z_{PIPE}$ : 12000 km/s. For both our method and $z_{PIPE}$, the large standard deviations are driven by the relatively large fraction of outliers, ie, catastrophic failures of redshift determination. The inter-quartile range for each method shows a measurement of dispersion which is not affected by these failures. We have: $\zmap$: 3,000 km/s $z_{VI}$ : 1,200 km/s, $z_{MgII}$ : 1,700 km/s, $z_{PIPE}$ : 1,600 km/s. Our redshift estimation method thus produces a larger dispersion than the other methods.

\begin{figure*}
   \includegraphics[width=2\columnwidth]{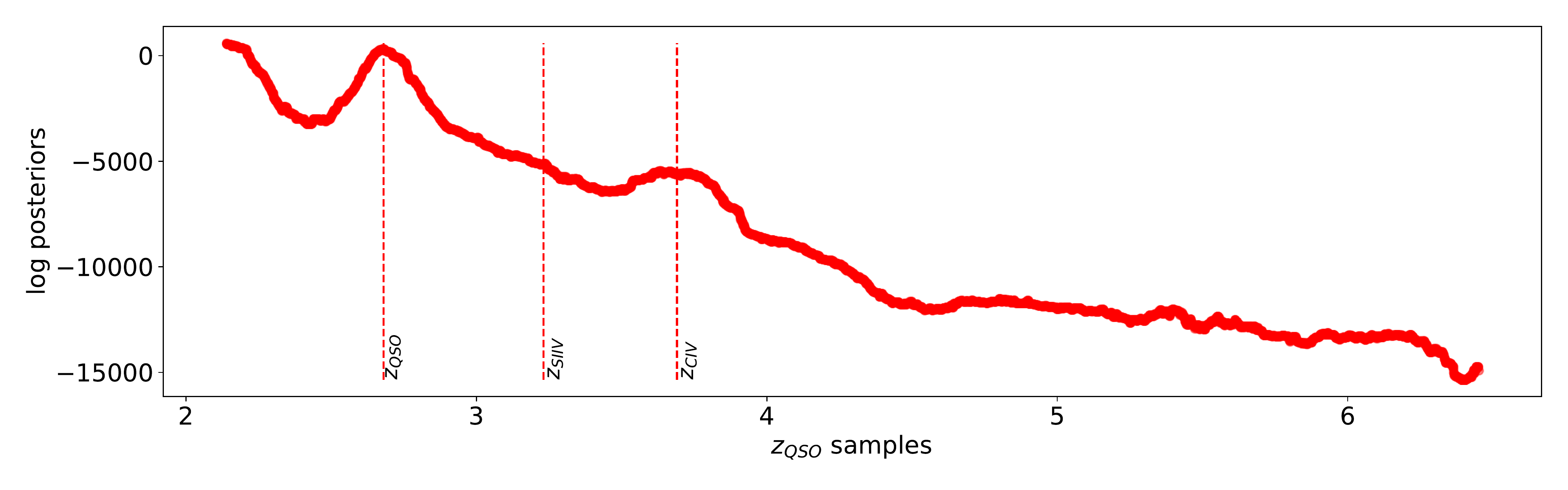}  \\
   \includegraphics[width=2\columnwidth]{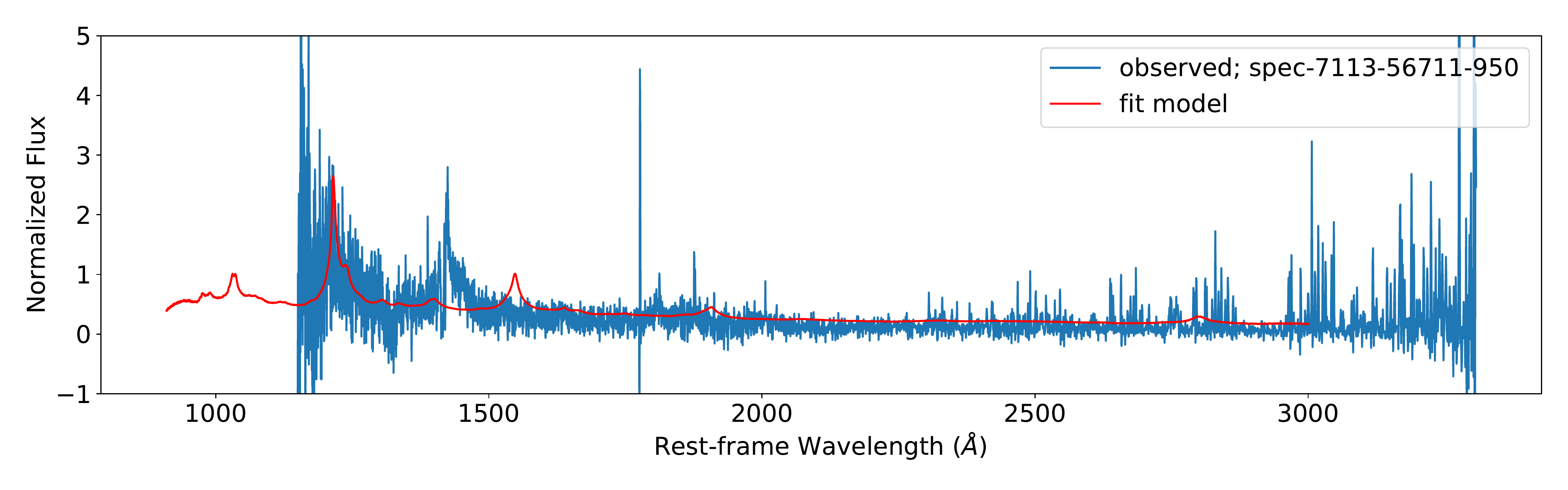} \\
   \includegraphics[width=2\columnwidth]{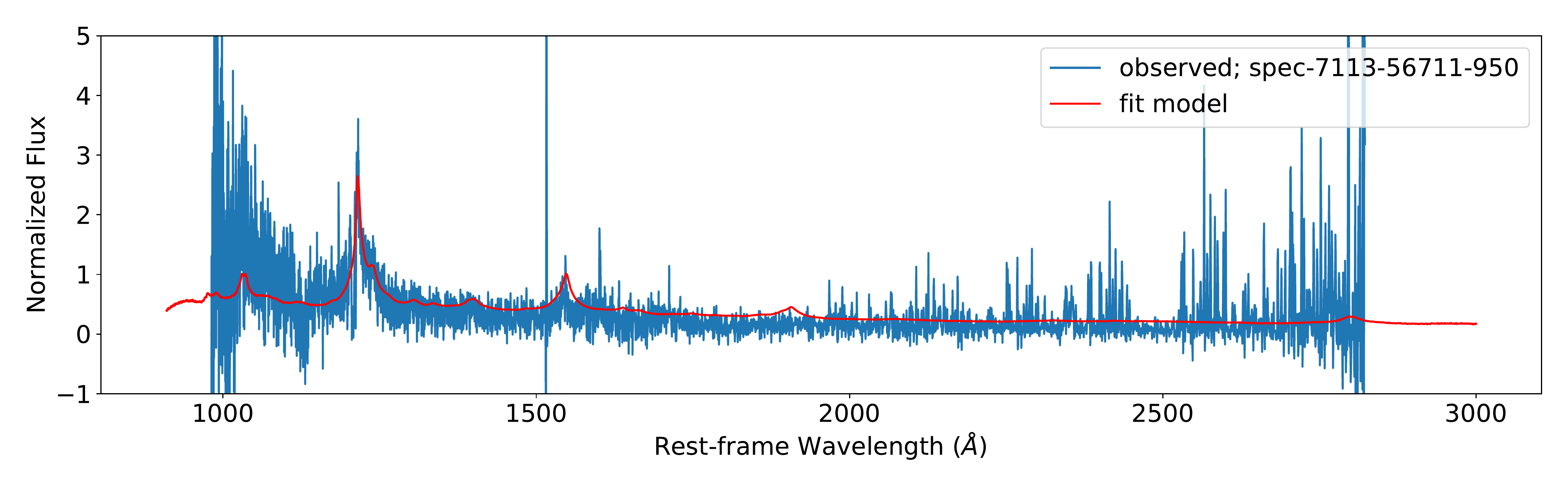}
   \caption{\textbf{(Top)} The sample posterior $p(\zqso \mid \obsflux, \mzqso)$ for a QSO with thingID $= 544031279$.
   The catalogue redshift is labelled as $\zqso$. Vertical dashed lines indicate the redshifts associated to samples at particular emission peaks. For example, the redshift resulting from trying to fit the true {\Lya} peak onto the observed CIV peak is shown as $z_\textrm{CIV}$. \textbf{(Middle)} The rest-frame spectrum using $\zmap$. \textbf{(Bottom)} The rest-frame spectrum using the SDSS visual inspection redshift $z_{VI}$. We use $z_{VI}$ as it is the method with the the lowest failure rate. The MAP value of our catalogue fits the \Lya~peak with what is really OVI.
   }
   \label{fig:bad_example_z_posterior}
\end{figure*}

We have visually inspected a subsample of the spectra where our catalogue has a dramatically incorrect redshift.
Figure~\ref{fig:bad_example_z_posterior} shows one such example. Here, the likelihood peaks at very low redshift, because the code
believes that a noise peak near the OVI emission line is the \Lya~peak, and this overwhelms the otherwise poor fit to the spectrum.
Note that there is a peak in the likelihood at the correct redshift, with almost the same probability, so a full Bayesian analysis
would be closer to the true value. This spectrum, like most of those where the code confuses OVI for \Lya, shows unusually noisy
data with an oscillatory feature which exceeds the expected pipeline noise at the far blue end of the observed data, possibly
related to the data reduction systematic identified by \cite{Lan:2018}. Spectra where the code confuses CIV for \Lya~often have unusually weak \Lya~peaks relative to their CIV emission.

\begin{figure*}
   \includegraphics[width=2\columnwidth]{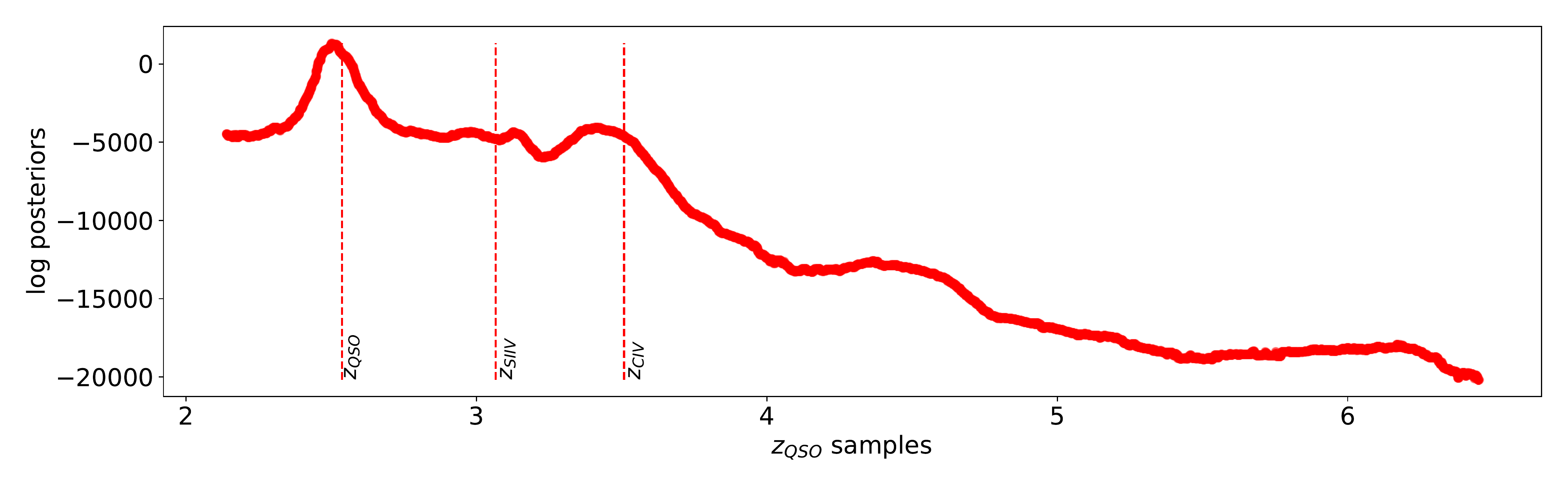}  \\
   \includegraphics[width=2\columnwidth]{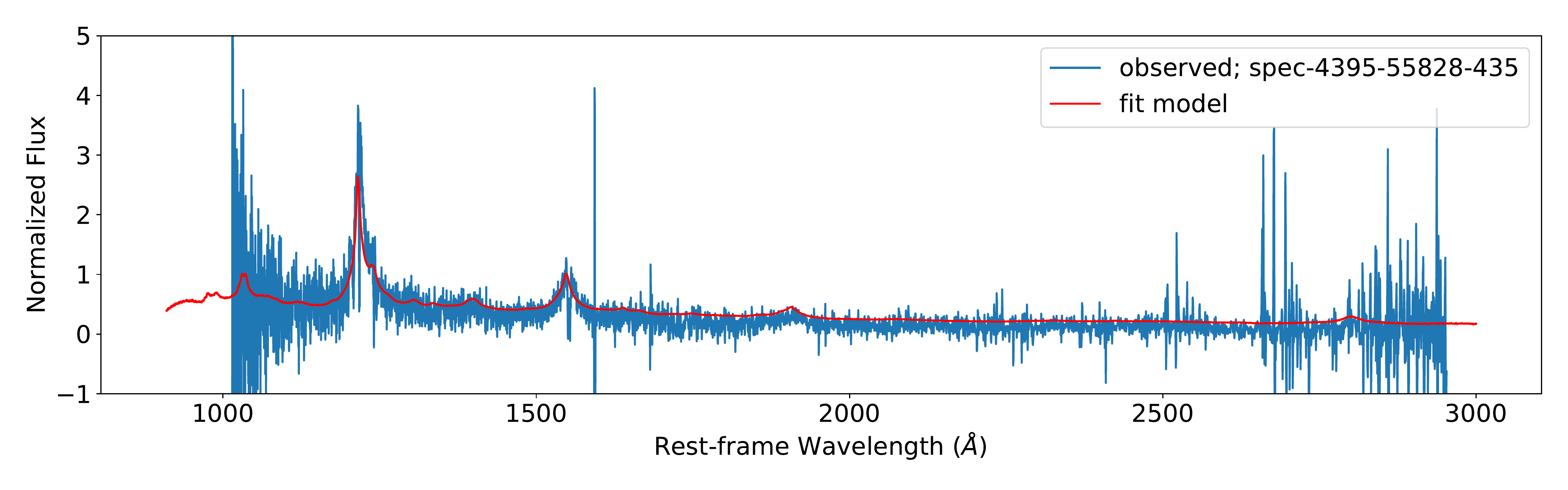} \\
   \includegraphics[width=2\columnwidth]{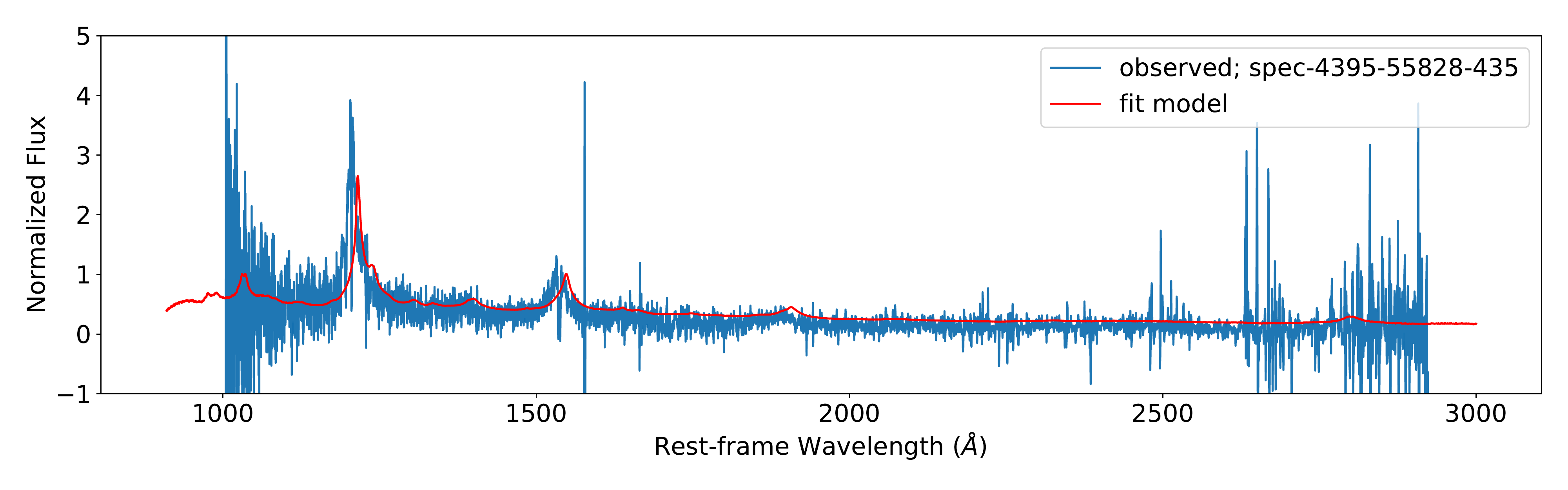}
   \caption{\textbf{(Top)} The sample posterior $p(\zqso \mid \obsflux, \mzqso)$ for a QSO with thingID $= 27885089$.
   The catalogue redshift is labelled as $\zqso$. Vertical dashed lines indicate the redshifts associated to samples at particular emission peaks. For example, the redshift resulting from trying to fit the true {\Lya} peak onto the observed CIV peak is shown as $z_\textrm{CIV}$. \textbf{(Middle)} The rest-frame spectrum using $\zmap$. \textbf{(Bottom)} The rest-frame spectrum using the SDSS visual inspection redshift $z_{VI}$. We use $z_{VI}$ as it is the method with the the lowest failure rate. $\zmap$ appears to produce a better fit than $z_{VI}$.}
   \label{fig:example_z_posterior}
\end{figure*}

There are also spectra in our catalogue where our method produces what looks visually like a better fit to the observed spectrum than $z_{VI}$. Figure~\ref{fig:example_z_posterior} shows an example, where the catalogue $z_{VI}$ redshift does not quite match the location of the CIV peak, possibly as an attempt to fit to noise near the MgII emission line. Our method estimates redshift as $\zqso = 2.501$. Redshift estimates from the SDSS catalogue are: $z_{VI} = 2.538$, $z_{PIPE} = 2.507$, $z_{PCA} = 2.511$. $z_{MgII}$ was not available. In this case $z_{VI}$ is an outlier, and our model is in reasonable agreement with $z_{PIPE}$. We note that the position of the CIV emission peak shown in the Figure is from the mean model, and thus automatically includes the average CIV blueshift from the rest-frame emission \citep{Hewett:2010, Richards:2011}.

\subsubsection{Validating the model at lower redshifts}
\label{subsec:low_z}
In this section, we validate the behaviour of our GP model $\mzqso$ on quasars with
redshift outside the redshift range containing DLAs. We place a uniform prior on $\zqso$ as in Eq.~\ref{eq:uniform_prior}, but we modify the lower bound to be $\zqsomin = 1.9$. We select the test set
as described in Section~\ref{sec:training} except that we modify the range of $\zqso$
to be $1.9 < \zqso < 2.15$. The new sample size is $16\,013$ quasars. We do not retrain the model.

The catastrophic misfit rate for $|z_{VI} - \zmap| > 0.5$ is $3.3\%$.
The error, as expected, is much larger than the results for spectra with $2.15 \leq \zqso$, as the \Lya~peak is now located at a lower observed frame wavelength, where instrumental noise is larger. Since we normalize by the height of the \Lya~peak, noise in this region can easily lead us to produce an inaccurate continuum. This normalization also leads to a natural minimum quasar redshift possible with our method at $\zqsomin = 1.9$, below which the \Lya~peak has not yet redshifted into the observation window of BOSS optical spectra ($3650-10400$~\AA).
We can achieve slightly improved results for lower $\zqso$ samples by using a GP model trained by normalizing on C\textsc{iv} peak, $1549 \pm 40 \AAtext$. Here the misfit rate was $2.8\%$ for $|z_{VI} - \zmap| > 0.5$. However, normalising to C\textsc{iv} performs substantially less well for quasars with $\zqso > 2.15$.


\subsection{DLA Finding}
\label{sec:dlaresult}

\begin{figure*}
\includegraphics[width=1.\textwidth]{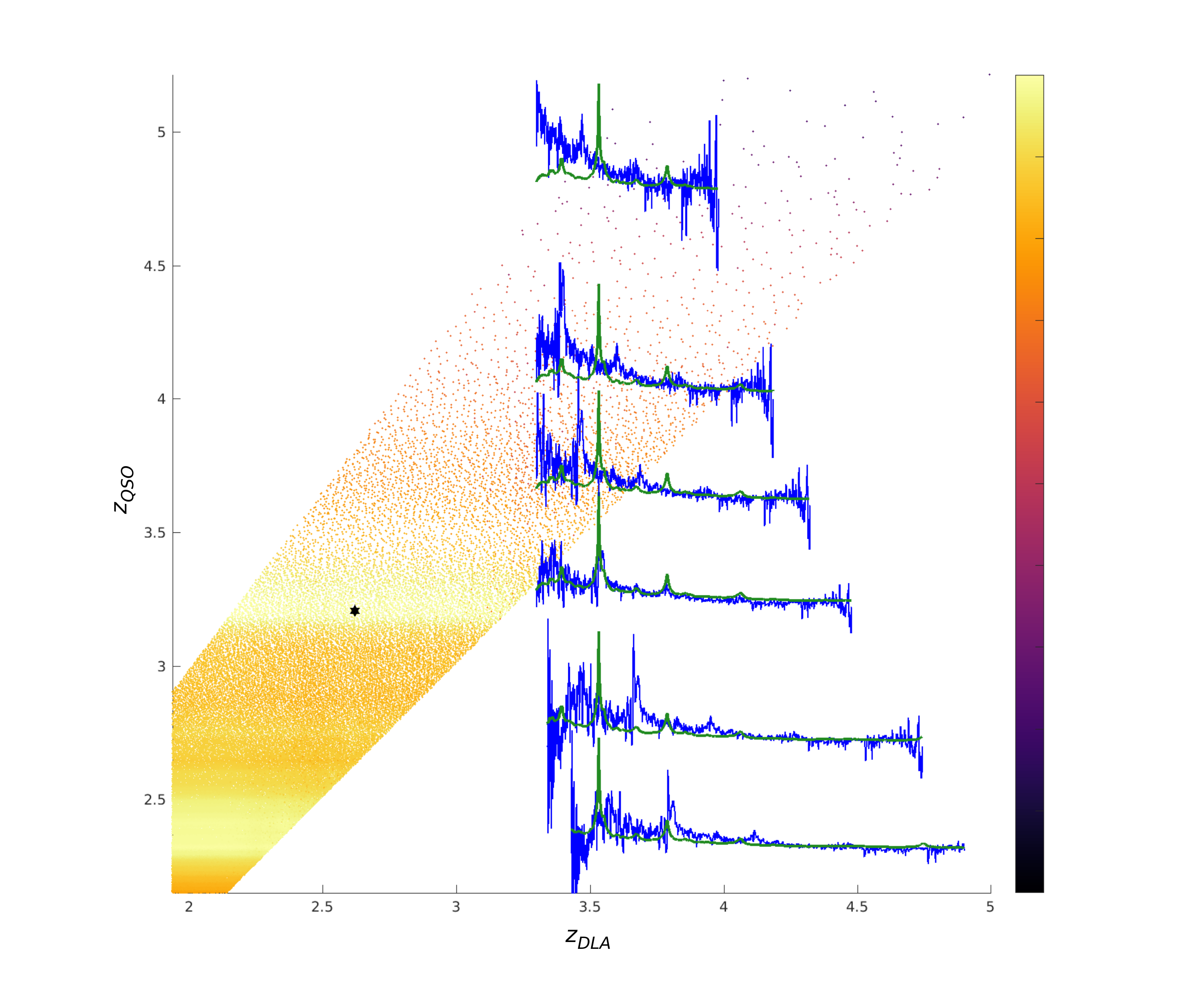}
\caption{Examples of Halton sequence sampling for $\zqso$, $\zdla$, and $N_{HI}$.
Samples across parameter space $\Theta$ project out $N_{HI}$ onto the $\zqso$ and $\zdla$ plane. The best sample (at $\zqso = 2.309$) is shown by a black star. Colours estimate the posterior log-likelihood of $\DLA$ for each point. $\zdla$ is drawn uniformly while $\zqso$ is taken from an empirical distribution. This particular quasar has a bimodal likelihood for $\zqso$, where the second, lower, peak corresponds to the code fitting the \Lya~peak at OVI. Though estimates of $\zdla$ are drawn uniformly, the DLA cannot appear redwards of the quasar or bluewards of the Lyman-$\alpha$ peak, and so are not sampled from these regions. Shown for reference in green are illustrations of the given quasar and rest-frame mean prediction in the rest-frame for $\zqso$ sampled at: $2.3$, $2.7$, $3.22$, $3.6$, $4.0$, $4.75$. Normalizations for the spectra are $1.93$, $1.43$, $1.80$, $1.06$, $0.78$, $0.51$, respectively.}
\label{fig:dlasamp}
\end{figure*}


We now show our DLA catalogue computed with a marginalized $\zqso$.
We have checked explicitly that redshift estimation is similar in this catalogue to the pure redshift estimation model discussed in Section~\ref{sec:zresult}.
A two-dimensional projection showing $\zqso$ and $\zdla$, for an example quasar with a DLA, can be seen in Figure~\ref{fig:dlasamp}. The mean over the product of each Bayes factor with each model prior for different $\zqso$ yields our posterior odds, which can be normalized to give our desired model posteriors $\Pr(\DLA | \mathcal{D})$ and $\Pr(\noDLA| \mathcal{D})$.

\subsubsection{Best 2/3 DLA Catalogue}

\begin{center}
\begin{figure}
\includegraphics[width=.45\textwidth]{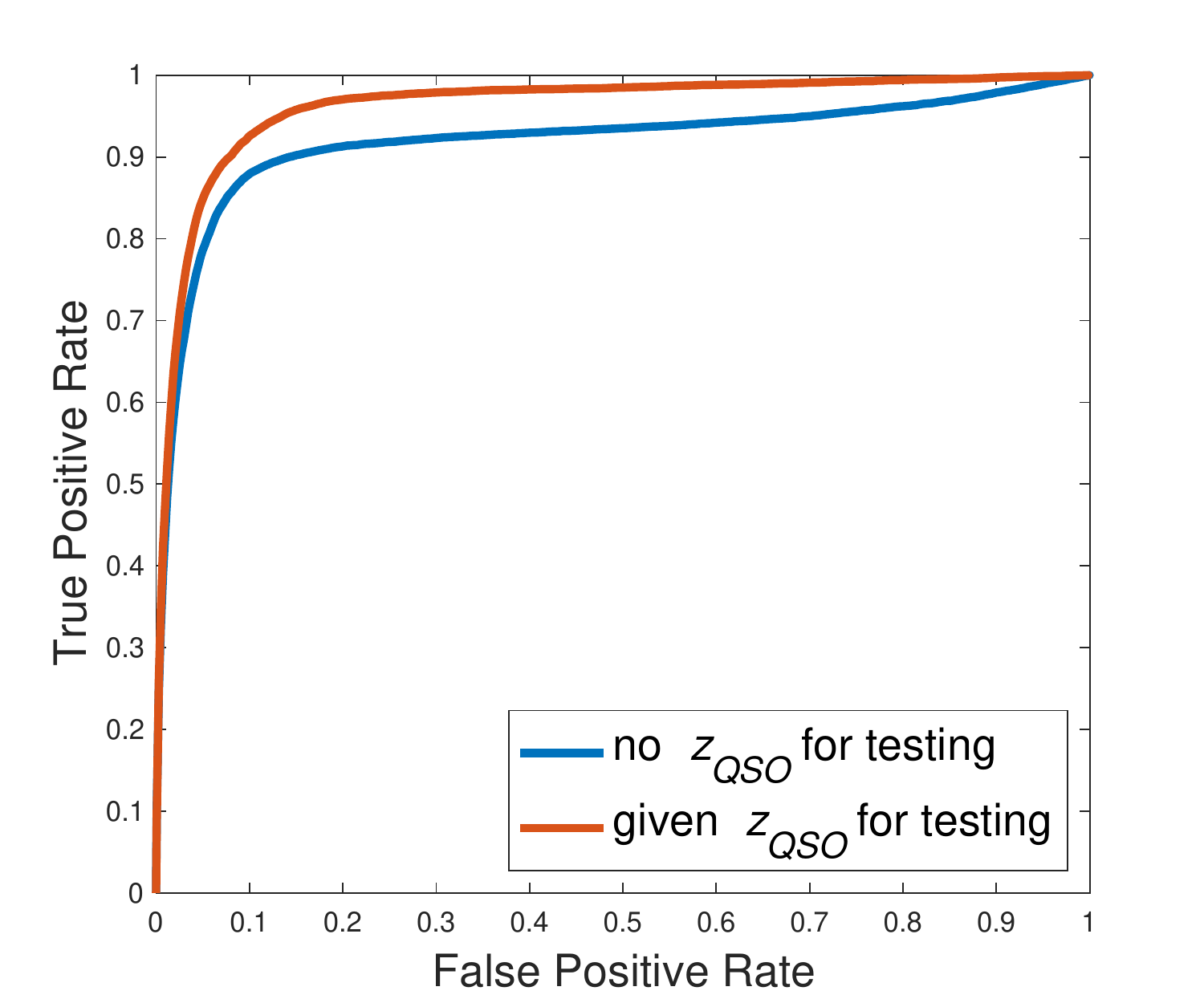}
\caption{ROC curve for DLA estimation from our catalogue estimating $\zqso$ (blue), and from the catalogue of \protect\cite{Ho:2020} with $\zqso$ given (red).
The AUC with full integration is $0.9192$. The AUC from \protect\cite{Ho:2020} is $0.9624$.
The ROC is taken over all $158,821$ applicable quasars in the DR12 dataset. Ground truth is the best 2-of-3 catalogue for DR12, described in the text.}
\label{fig:auc}
\end{figure}
\end{center}

%
To compare our results to a single ``ground truth'' DLA catalogue, we follow a procedure similar to that used to generate the DR9
concordance DLA catalogue \citep{lee2013boss}. Aside from our work, there are three extant DR12 catalogues. These are \cite{parks2018deep}\footnote{We include subdlas from this catalogue so that the minimum column density from all catalogues is $10^{20}$ cm$^{-2}$, as the other catalogues.} (based on a neural network), a DR12 catalogue generated using the template matching method of \cite{noterdaeme2012column} and the DR12 visual survey \citep{paris2017sloan}.\footnote{All DLAs of which we assign an arbitrary column density of $2\times 10^{20}$ cm$^{-2}$.} Each method produces a slightly different DLA catalogue, differing by up to $\sim 10\%$. However, by taking only DLAs which occur in 2/3 catalogues, we hope to produce a relatively pure sample.

To demonstrate our model effectiveness, we order each spectrum by its log posterior odds of $D^1$, with associated DLA information. Spectra which are assigned a DLA by our best 2/3 catalogue should appear at the top of this ordering as most probable. Figure~\ref{fig:auc} shows the receiver-operating characteristic (ROC) plot of each method, comparing our current method integrating over $\zqso$ to a model with $\zqso$ assumed known \citep{Ho:2020}. The AUC between our $\zqso$ marginalizing catalogue with full $\zqso$ integration and the best 2/3 is $0.9192$. The AUC with known redshifts is $0.9624$. The AUC between our current catalogue and that with known redshifts was $0.914$, similar to the AUC between the $\zqso$ catalogue and the best 2/3.

Our method performs moderately less well than a similar integration task where $\zqso$ is given. This is not surprising, as the integration task without $\zqso$ is more difficult. While both models ultimately recover similar information, the full integration method estimates DLAs with less certainty, leading to a true positive rate which is worse by a few percent. When a DLA is correctly identified the MAP DLA redshift and column density is similar to our previous papers, exhibiting no noticeable preference for higher or lower column densities. In particular, there are several instances where the DLA redshift is correctly determined despite the quasar redshift being incorrect.\footnote{This is possible because the transformation between observed frame and DLA frame does not depend on the quasar rest-frame, as long as the measured $\zqso$ allows for a DLA in the observed region.}



\begin{figure}
\includegraphics[width=.5\textwidth]{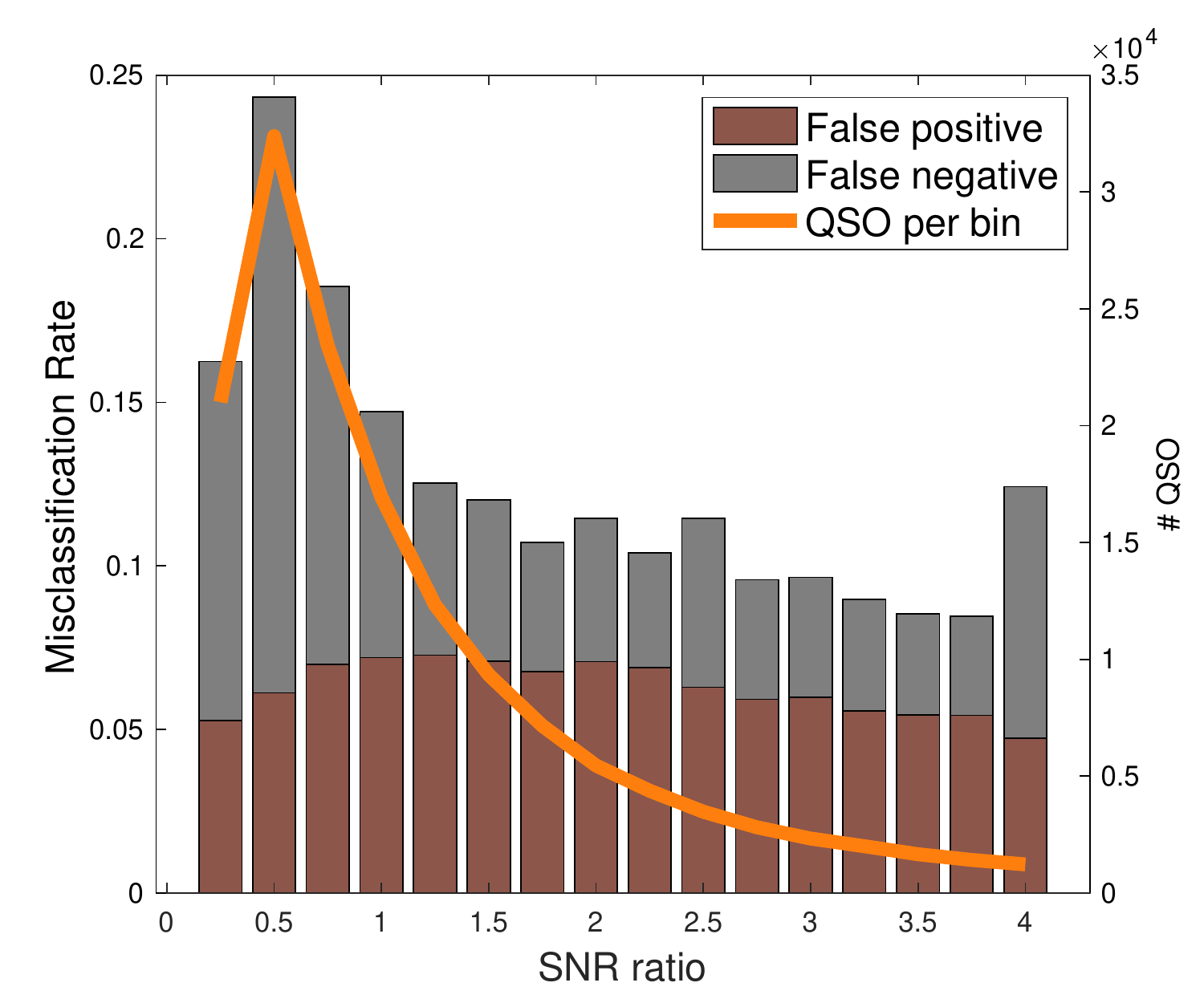}
\caption{Error rate plotted as a function of signal-to-noise ratio. Curve and right y-axis shows the total number of quasars in each signal-to-noise bin, while the left y-axis shows the error rate. We consider that a spectrum has a DLA in our catalogue if $p(DLA) > 0.9$. Low SNR spectra have a higher level of false negatives. Ground truth is the best 2-of-3 catalogue for DR12, described in the text.}
\label{fig:stn}
\end{figure}

If our lower true positive rate is due simply to the increased difficulty of the
problem, the presence of spectral noise should reduce the ability of our model to
determine $\zqso$. Figure~\ref{fig:stn} shows the error rate as a function of our
catalogue's signal-to-noise ratio. Signal-to-noise was taken over as much
of each quasar as could possibly sit in the rest-frame, as a per-pixel mean of the
flux over the square root of the noise variance. Also shown is the overall frequency
of quasars per bin. Our false negative rate is indeed higher by a factor of two at
low SNR. This may indicate that false negatives occur because there is not enough
information for the model to make a solid detection. It is also possible that that these are not, in fact, real DLAs, and the
low signal-to-noise ratio was causing a slightly incorrect pipeline $\zqso$ which was misleading our previous DLA algorithm.

We have visually inspected a sample of low signal-to-noise spectra with false positive DLAs and poor redshift estimation. There are
several examples where only $0$--$1$ emission peaks emerge from the noise. Our false positives commonly occur in spectra where, if one takes the SDSS pipeline redshift as ground truth, one observes a Lyman break with noise at $700-800$~\AA. Our pipeline instead fits the OVI emission peak with \Lya~and interprets the break as a DLA. We suspect that most of these cases are indeed false positives, but obtaining reliable results from SNR $< 1 $ will always be challenging.

\begin{figure}
\includegraphics[width=.5\textwidth]{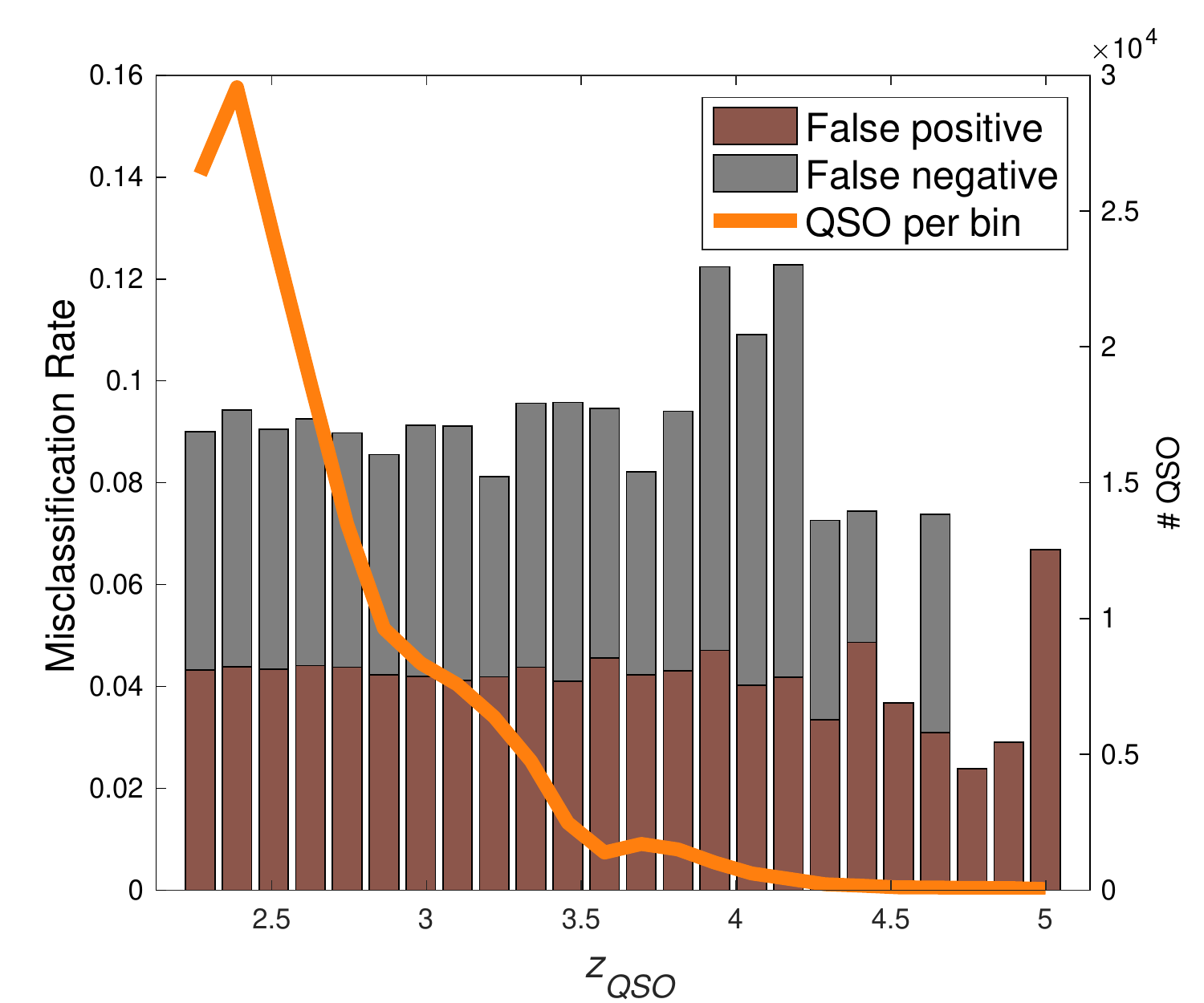}
\caption{Error rate plotted as a function of quasar redshift. Curve and right y-axis shows the total number of quasars in each redshift bin, while the left y-axis shows the error rate. We consider that a spectrum has a DLA in our catalogue if $p(DLA) > 0.9$. Ground truth is the best 2-of-3 catalogue for DR12, described in the text.}
\label{fig:zfp}
\end{figure}

Figure~\ref{fig:zfp} shows the error rate as a function of quasar redshift. The false positive rate is roughly independent of
redshift, while the false negative rate is constant until $z=3.6$. At $z>4.2$ the false negative rate approaches zero. However,
there are very few DLAs detected at this redshift in the best $2/3$ catalogue. For $z=3.7$ -- $4.0$ the false negative rate increases noticeably. In this redshift range the Lyman break at $910$~\AA~redshifts into the observed SDSS band, and it may be that our redshift estimation was confused by the presence of this feature in the spectrum.

\section{Conclusion}
\label{sec:conclusion}

We have extended our Gaussian process based code for finding DLAs in SDSS quasars to situations where the quasar redshift is not
known. This required extending the Gaussian process range to encompass more emission lines and thus get a more reliable $\zqso$ estimate. It was also necessary to augment the model to include a likelihood component for all observations, even those which are outside the range of the Gaussian process, so that the probabilities are comparable for the same spectrum across multiple redshifts.

We first estimated the redshift of the SDSS DR12 sample, showing that our redshift labelling is competitive to existing
redshift estimation. Large redshift misestimation was reasonably rare. Our redshift estimate differs from the PCA redshift by $> 0.5$ for $603$ quasars out of $\sim 1.6\times 10^5$. The median redshift error of our method compared to other SDSS redshift estimates was $\sim 100$ km/s.
We used our improved model to find DLAs while marginalizing over uncertainties in the quasar redshift. We detected a few percent fewer DLAs at high confidence than our earlier methods (AUC drops from $0.96$ to $0.91$) , especially in noisy spectra where estimation is more difficult.

The computation time for the pure redshift estimation model is $\sim 1.5$ seconds per spectrum on a 48-core AWS EC2 machine, while finding DLAs takes $\sim 60$ seconds per quasar.

There are a few ways in which the redshift estimation present here may be improved. Our choice of normalization (the \Lya~peak) makes low redshift quasars hard to classify correctly. In future work it might be better to incorporate normalization directly into the Bayesian model as an extra parameter. We may also have reached the limits of the Halton sequence based quasi Monte-Carlo integrator we have used since \cite{Garnett:2016}. Future work may find it necessary to switch to a
more targeted integrator based on variational or Markov chain Monte Carlo methods.

\section*{Acknowledgements}

We thank Yongda Zhu and Marie Wingyee Lau for useful conversations. SB was supported by NSF grant AST-1817256. RG was supported by the NSF under award
numbers IIS--1939677, OAC--1940224, and IIS--1845434. SB and RG were supported by an Amazon.com Machine Learning Research Award, which also provided computing time. CS was supported in part by NSF grant (IIS 1510741). Computing time was also provided by UCR HPCC.

\section*{Data availability}
All the code to reproduce the data products is available in our GitHub repo: \url{https://github.com/sbird/gp_qso_redshift}.
The final data products are available in this Google Drive: \url{http://tiny.cc/gp_zestimation_catalogue}, including a MAT (HDF5) catalogue and a JSON catalogue.

\label{lastpage}
\bibliography{paper}

\end{document}